\documentclass{aa} 
\usepackage[T1]{fontenc} 
\usepackage{siunitx}
\usepackage{amsmath}
\usepackage{amssymb}
\usepackage{graphicx}
\usepackage{longtable} 
\usepackage{lscape}
\usepackage[export]{adjustbox}
\usepackage{natbib}
\usepackage{subcaption}
\usepackage{calrsfs}
\usepackage{array}
\usepackage{hyperref}
\usepackage{multirow}
\usepackage{gensymb}
\usepackage[capitalise]{cleveref} 

\bibpunct{(}{)}{;}{a}{}{,}

\newcommand{\Mjup}{M$_{\rm Jup}$\,}
\newcommand{\Mjupv}{M$_{\rm Jup}$}

\newcommand{\Msun}{M$_{\sun}$}

\newcommand{\msini}{$m_{\rm p}\sin{i}$\,}

\begin{document} 

\title{On the radial distribution of giant exoplanets at Solar System scales}


 \author{A.-M. Lagrange\inst{1,} \inst{2}\fnmsep\thanks{Please send any request to Anne-marie.Lagrange@obspm.fr}
 \and
 F. Philipot\inst{1} 
 \and
 P. Rubini\inst{3} 
 \and
 N. Meunier\inst{2} 
 \and
 F. Kiefer\inst{1} 
 \and
 P. Kervella\inst{1} 
 \and
 P. Delorme\inst{2} 
 \and
 H. Beust\inst{2}
 }

 \institute{
LESIA, Observatoire de Paris, Universit\'{e} PSL, CNRS, 5 Place Jules Janssen, 92190 Meudon, France
\and
Univ. Grenoble Alpes, CNRS, IPAG, 38000 Grenoble, France
\and
Pixyl S.A. La Tronche, France
 }
 \date{Received XXX / Accepted XXX}

 
 \abstract
 {Giant planets play a major role in multiple planetary systems. Knowing their demographics is important to test their overall impact on planetary systems formation. It is also important to test their formation processes. Recently, three radial velocity surveys have established radial distributions of giant planets. All show a steep increase up to 1-3 au, and two suggest a decrease beyond.}
 {We aim at understanding the limitations associated with the characterization of long-period giant radial velocity 
 planets, and to estimate their impact on the radial distribution of these planets.}
 {We revisit the results obtained by two major surveys that derived such radial distributions, using the RV data available at the time of the surveys as well as, whenever possible, new data.}
 {We show that the radial distributions published beyond (5-8 au) are not secure. More precisely, the decrease of the radial distribution beyond the peak at 1-3 au is not confirmed.}
 {}


 \maketitle

\section{Introduction}

Our giant planets played a significant role in the building of the Solar System (\cite{2003AJ....125.2692L}, \cite{2014prpl.conf..595R}, \cite{2012AREPS..40..251M}, \cite{2020PASP..132j2001H}), and, possibly, also in the development of life on Earth \citep{2020AJ....159...10H}. Extra-solar giants are also thought to play a key role in the formation and evolution of planetary systems (\cite{2020PASP..132j2001H}, \cite{2016AAS...22840405Q}, \cite{2019MNRAS.485..541C}), and, in particular, in the presence or absence of rocky planets in their Habitable Zones \citep{2003AJ....125.2692L}. From an observational point of view, giant planets may hinder the detection of lighter and closer-in planets with indirect techniques such as radial velocity (hereafter RV) and astrometry.
Last, the radial distribution of giant planets can give precious constraints on the formation processes and early evolution when compared to the outputs of population synthesis models (\cite{2018haex.bookE.143M}, \cite{2021A&A...656A..70E}, \cite{2021A&A...656A..73S}, \cite{2021A&A...656A..72B}, \cite{2021A&A...656A..71S}). For these reasons, it is important to get an accurate view of their demographics as well as of individual systems.

Indirect methods (transit photometry, RV, astrometry) are, in principle, best suited to determine the radial distribution of mature giant planets (\cite{2018exha.book.....P}, \cite{2021exbi.book....2G}) within 10 au. While transit photometry is appropriate for finding exoplanets below $\simeq $ 2 au, RV can perform long monitoring, precise enough to find more remote, long period (decades) giant planets around solar-type stars \citep{2021exbi.book....2G}. Finally, while siblings of Jupiter (similar distance, similar mass, similar age) may be within reach of high contrast imagers mounted on forthcoming ELTs, through reflected light (and may be also through thermal emission), direct detection is best suited today to detect the thermal emission of young planets beyond 10 au from their host stars (\cite{2010MNRAS.407...71V}, \cite{2019AJ....158...13N}).

\begin{figure}[t!]
 \centering
\includegraphics[width=0.5\textwidth]{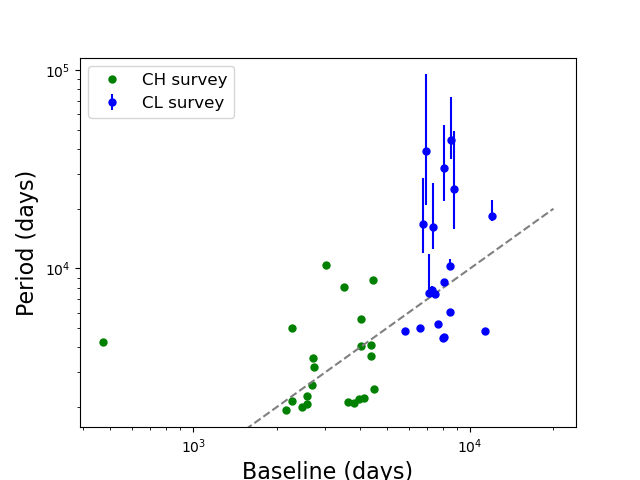}
\caption{Orbital periods vs temporal baselines for reported planets with \textit{a} > 3 au (CH survey) and \textit{a} > 5 au (CL survey). The uncertainties are indicated, whenever available. The dotted line represents the limit where the period is equal to the baseline. 
\label{baseline_period}} 
\end{figure}

Three large spectroscopic surveys have estimated the radial distribution of giant planets orbiting up to 10-30 au from solar-type stars:

\begin{itemize}

 \item The CORALIE-HARPS (hereafter CH) survey \citep{2011arXiv1109.2497M} monitored 822 stars and found 155 planets with semi-major axis (hereafter \textit{a}) up to 9.4 au. Based on the detection and an estimate of the survey completeness given in the CH survey, the radial density distribution of giant planets with masses of 30-6000 $M_{Earth}$ up to more than 10 au is estimated in a recent study (hereafter referred to as CHS; \cite{2019ApJ...874...81F}). The distribution is shown to increase up to 2-3 au, and then to decrease beyond. This “so-called” turnover is well reproduced by population synthesis models \citep{2018haex.bookE.143M} coupling the core accretion scenario \citep{2008ApJ...673..502K} and migration theories. This result has since been used in the literature to validate ab initio modeling of planetary formation and early evolution processes that predict giant planets' distribution rates (\cite{2021A&A...650A.116M}, \cite{2021AJ....162...28V}, \cite{2022ASSL..466....3R}), or to propose observing strategies with forthcoming instruments \citep{2020ApJ...893..122D}. However, the temporal coverage of the RV variations induced by the long-period giant planets orbiting beyond 3-5 au is rather poor (see Figure \ref{baseline_period}), which may have an impact on the orbital parameters and minimum masses estimates. In fact, since the CH survey, the orbital parameters of 6 targets have already been significantly revised, thanks to additional data, and 2 new planets have been identified.
 \item The Anglo Australian Planetary survey (hereafter AAPS; \cite{2020MNRAS.492..377W}) monitored 203 stars with the Anglo Australian Telescope\footnote{This survey also used, when possible, HARPS, CORALIE and HIRES data.}, and reported 38 planets with periods of up to 10000 d. Despite the limited sample size, the analysis suggested a steep increase in the density of giant planets (defined as with masses in the range 0.3-13 \Mjupv ) orbiting up to 1 au from the star, followed by a possible flat distribution.
 \item The California Legacy (hereafter CL) survey \citep{2021ApJS..255....8R} included 719 F-G-K-M stars\footnote{Note that the sample excluded known metal-rich stars, and some other targets that could induce possible biases.} monitored over more than 20 years, and reported 177 giants with 0.1 < \textit{a} < 45 au and 0.1 < \msini < 47 \Mjup (uncertainties included). Based on the detection and on an estimate of the survey completeness, a radial distribution is estimated for giants with masses of 30-6000 $M_{Earth}$, and semi-major axes up to more than 30 au. This study (hereafter CLS, \cite{2021ApJS..255...14F}) reported a steep increase of the giant planets occurrence rate between 1 and 3 au, followed by a relative plateau for \textit{a} up to 7-8 au (period about 8000 d), and a possible fall-off at larger orbital distances, a scenario “favored over models with flat or increasing occurrence”. However, the temporal coverage of the long-period giant planets with \textit{a} > 7-8 au is poor (see Figure \ref{baseline_period}), which, as in the case of the CH survey, should call into question the derived conclusions.

\end{itemize}

Even though the CH and CL analyses describe the decrease of the radial distribution beyond the observed peak as tentative/possible, many studies now assume these radial distributions as valid at all separations. It is therefore important to have a deeper look at the characterization of these (very) long-period planets, and their radial distribution. 

The detection and characterization of long-period RV planets is a complicated task. An illustrative example is HD 7449Ac, for which the CH survey, using data spread over 4000 d and a genetic algorithm to fit the data, reported a period of 4046 days and a minimum mass of 2 \Mjupv. A subsequent study \citep{2019MNRAS.484.5859W}, based on data spread over 4452 days, and using MCMC fitting, instead reported a period of 15441 ± 1059 days (\textit{a} = $12.7 \pm 0.6$ au) and a minimum mass of $19.2 \pm 4.2$ \Mjupv. Independently, high-resolution images (\cite{2016ApJ...818..106R}, \cite{2017A&A...602A..87M}) revealed the companion at a projected separation of about 20 au in 2015-2016, and showed that it is in fact a 0.17 \Msun mass star, orbiting at $\sim $ 18 au from HD 7449A. This motivated the present study, which aims to i/ analyze the difficulties of the characterization of long period giant planets using RV data, ii/ revise the orbital parameters/minimum masses of those identified in the CH and CL surveys, and iii/ reconsider the radial distribution of such planets.

In this paper, we, therefore, investigate several factors that impact the detectability of exoplanets (Section 2). Then, we describe our approach to analyze the RV data of the stars considered in \cite{2019ApJ...874...81F} and \cite{2021ApJS..255...14F} (Section 3). The results of this new analysis are shown in Section 4. The implication on the radial distribution of long-period giant planets is provided in Section 5. Finally, using the data collected by the CL survey, we test the detectability of Solar System analogs with RV technique (section 6) and conclude in Section 7.

\section{Detection and characterization of long-period giant planets with RV data: issues}

The ability to detect and accurately characterize RV long-period giant planets is limited when the time coverage of the RV variations (\cite{2005Natur.434..873F}, \cite{2022ASSL..466..237G}) is poor (limited time baselines, uneven temporal samplings). This is true, in particular, when the RV time-series do not cover a maximum and/or a minimum of the RV signature of the planet. The poor orbital characterizations are due to the presence of noise in the data (stellar noise, instrumental noise), to the unknown star’s RV, or to a combination of these effects.

\subsection{Stellar activity/noise}

Stellar activity occurs on different timescales \citep{2021arXiv210406072M}, ranging from minutes (solar-type oscillations) to hours (granulation), days (supergranulation), days/weeks (spots, faculae) and years/decades (magnetic cycles). Short-term variations are usually accounted for by jitter in the fitting procedures. Magnetic cycles can lead to false detections, or poor accuracy of planet parameters, if the planets have a similar period (and phase) to the magnetic cycle (\cite{2013ApJ...774..147R}, \cite{2016ApJ...818...34E}). To take this long-term activity into account, one should ideally fit simultaneously the activity and planetary RV signals using spectroscopic tracers of magnetic activity. An alternative is to correct the RV signal from the stellar activity signal as estimated from such proxies. This approach, which was adopted in the CH survey, using the measured log(R’HK), may however also partially remove the planet’s induced RV variations as well (\cite{2019A&A...632A..81M}, \cite{2021MNRAS.505..830C}). Another approach was used in the CL survey, where potential planets giving signatures in the periodograms close to identified activity peaks were rejected. This might lead to an underestimation of the number of detected planets in the corresponding semi-major axes range (3-5 au). Furthermore, if not removed, the activity signal also may also impact the detection and characterization of planets with periods different from the cycle length. 

To evaluate the impact of an uncorrected long-term activity cycle on the characterization of the long-period giant planets (hereafter LPGP) identified with \textit{a} > 3 au (CH survey) or \textit{a} > 5 au (CL survey), we estimate the magnetic activity signal of each star, using the results of earlier simulations of stellar activity (\cite{2019A&A...632A..81M},\cite{2019A&A...625L...6M})\footnote{Note that using spectroscopic activity tracers for each star is not possible in the present study as, in most cases, such data are not available.}. The approach is described in Appendix A. The amplitudes associated with the activity signal over the calendar of observations are found to be between 2 and 15 m/s. For each star, we compute the most impacted domain of (\textit{a}, \msini) (see Figure \ref{sma_mass_activity}). It turns out that the stellar activity signal will mainly impact the detection and characterization of planets < 0.6 \Mjup orbiting beyond typically 3 au. More massive planets at similar separations produce higher amplitude RV signals; their characterization will be impacted to a lesser extent. Similarly, higher-mass planets orbiting further away could be impacted but to a lesser extent.

\subsection{Radial velocity of the star}

\begin{figure}[t!]
 \centering
\includegraphics[width=0.43\textwidth]{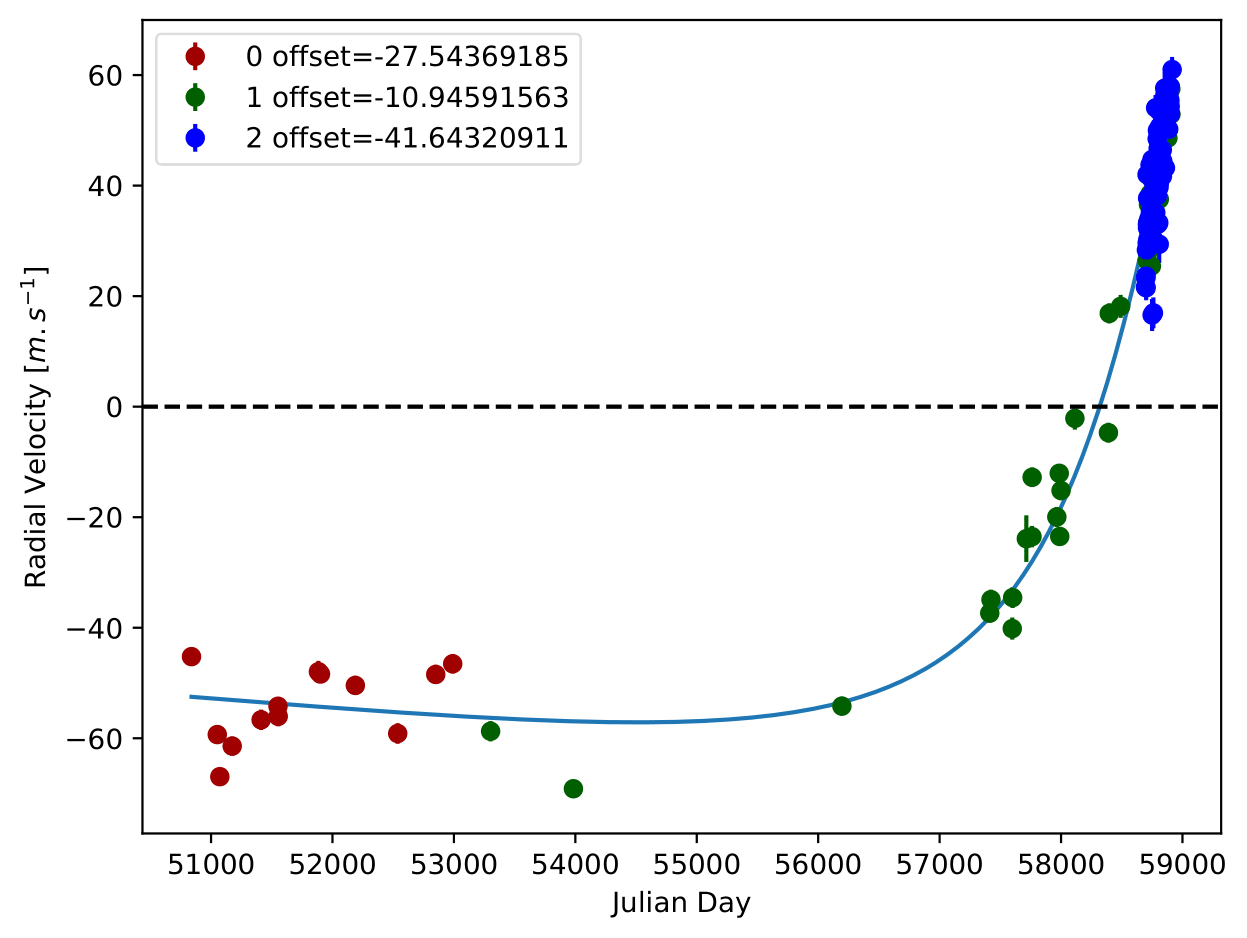}
\includegraphics[width=0.43\textwidth]{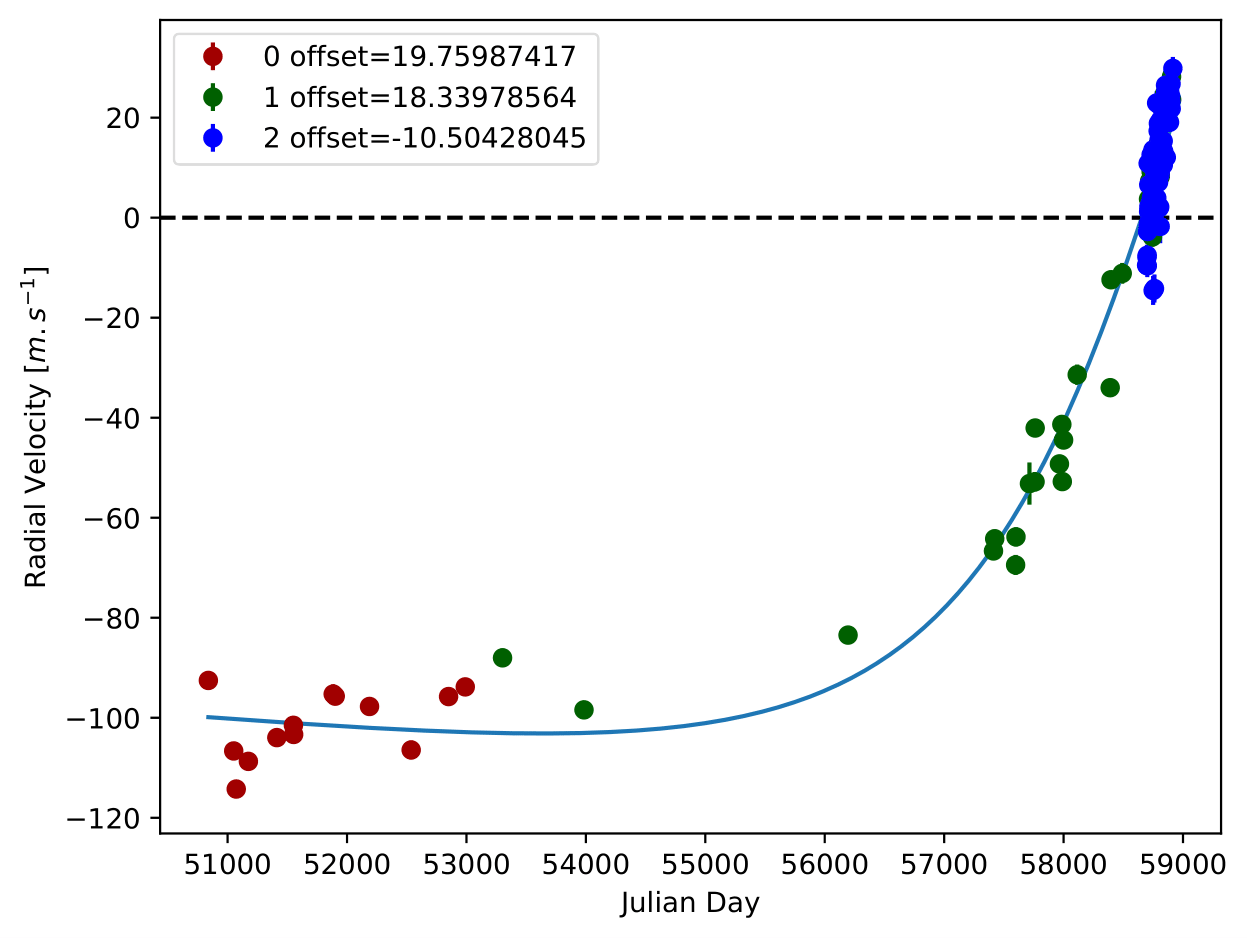}
\includegraphics[width=0.43\textwidth]{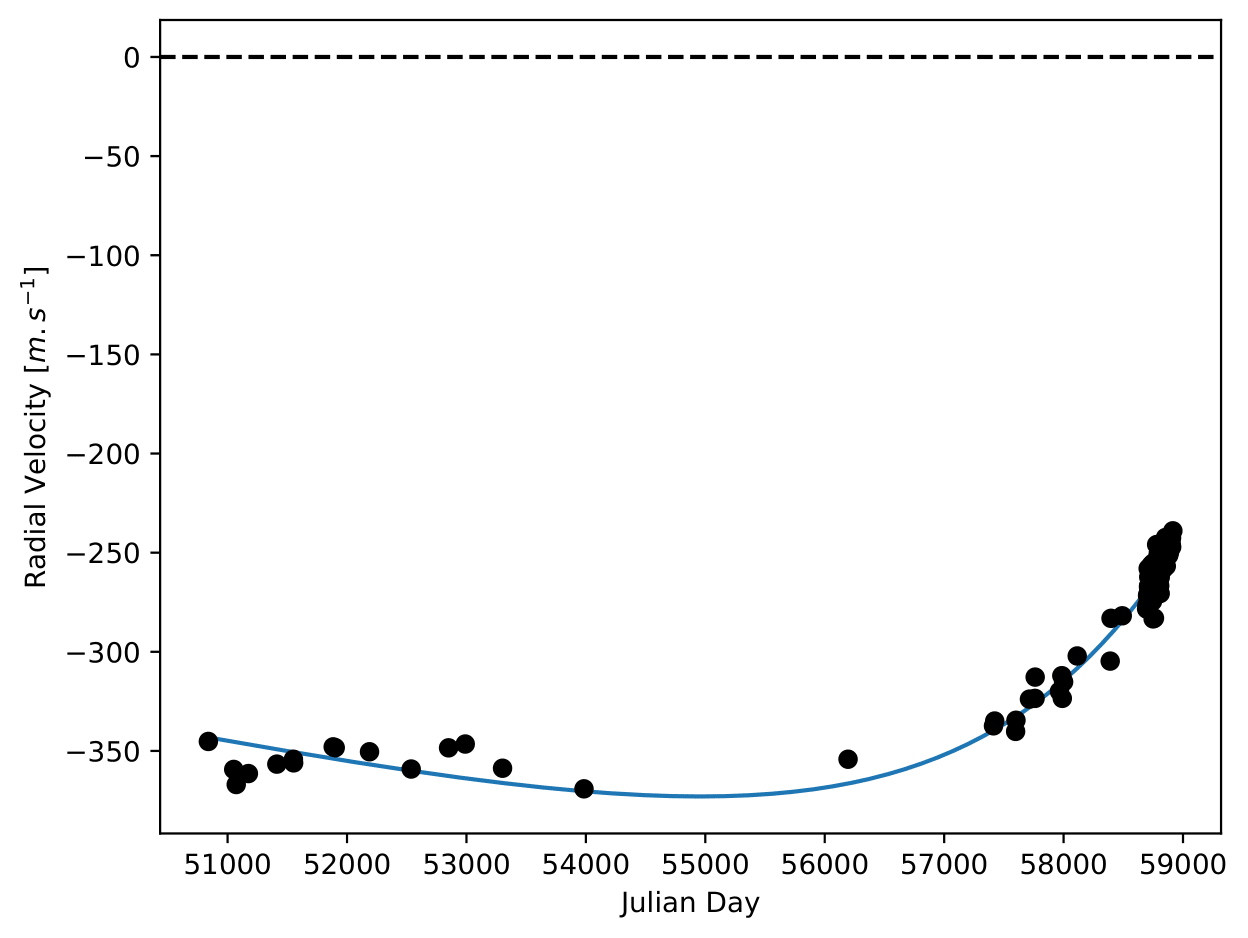}
\caption{Degeneracies between the star RV and the planet orbital parameters and masses. Case of HD 26161. RV time series (filled circles) and fit with a genetic algorithm (blue curve). \textit{Top}: All priors are left free. Red – Hir94; green – Hir04; blue – APF. \textit{Middle}: The semi-major axis is fixed to 240 au. Same color codes as on the \textit{top}. \textit{Bottom}: The star RV is fixed to 300 m/s. The black points correspond to RVs corrected from the instrumental offsets for clarity purposes. The semi-major axes found are respectively 33 au, 240 au, and 42 au and the minimum mass found are respectively 14 \Mjupv, 23 \Mjupv, and 80 \Mjup while the rms of the residuals are almost identical (respectively 6.2 m/s, 7 m/s, and 7.2 m/s). 
\label{ex_HD 26161}} 
\end{figure}

Temporal coverage of the planet-induced RV variations shorter than typically half of the planet’s period leads to degeneracies between the planet's properties and the star’s intrinsic RV (a constant value for a single star, which acts as an offset). This is particularly true when only one extremum or even none of the extrema is covered. Such degeneracies are not or are only partly captured by the fitting tools used (see Section 3.2). An extreme example is shown in Figure \ref{ex_HD 26161}, in the case of HD 26161, whose RV time series includes neither a minimum nor a maximum. The CL survey reported a companion with a period of $32000_{-10000}^{+21000}$ days (\textit{a} of $20.4_{-4.9}^{+7.9}$ au), i.e. 4 times larger than the time baseline (8706 d), and a minimum mass of $13.5_{-3.7}^{+8.5}$ \Mjupv. Shifting the RV of the star by 300 m/s (i.e. 2.5 times the uncertainty associated with the RV measured by Gaia over six years) leads to significantly different orbital parameters and minimum mass: 42 au and 80 \Mjupv. More generally, assuming various values of star RV offsets leads to a much wider range of orbital and mass solutions (see Section 3 for a detailed description of the procedure).

\subsection{Instrumental RV offsets}

\begin{figure}[t!]
 \centering
\includegraphics[width=0.45\textwidth]{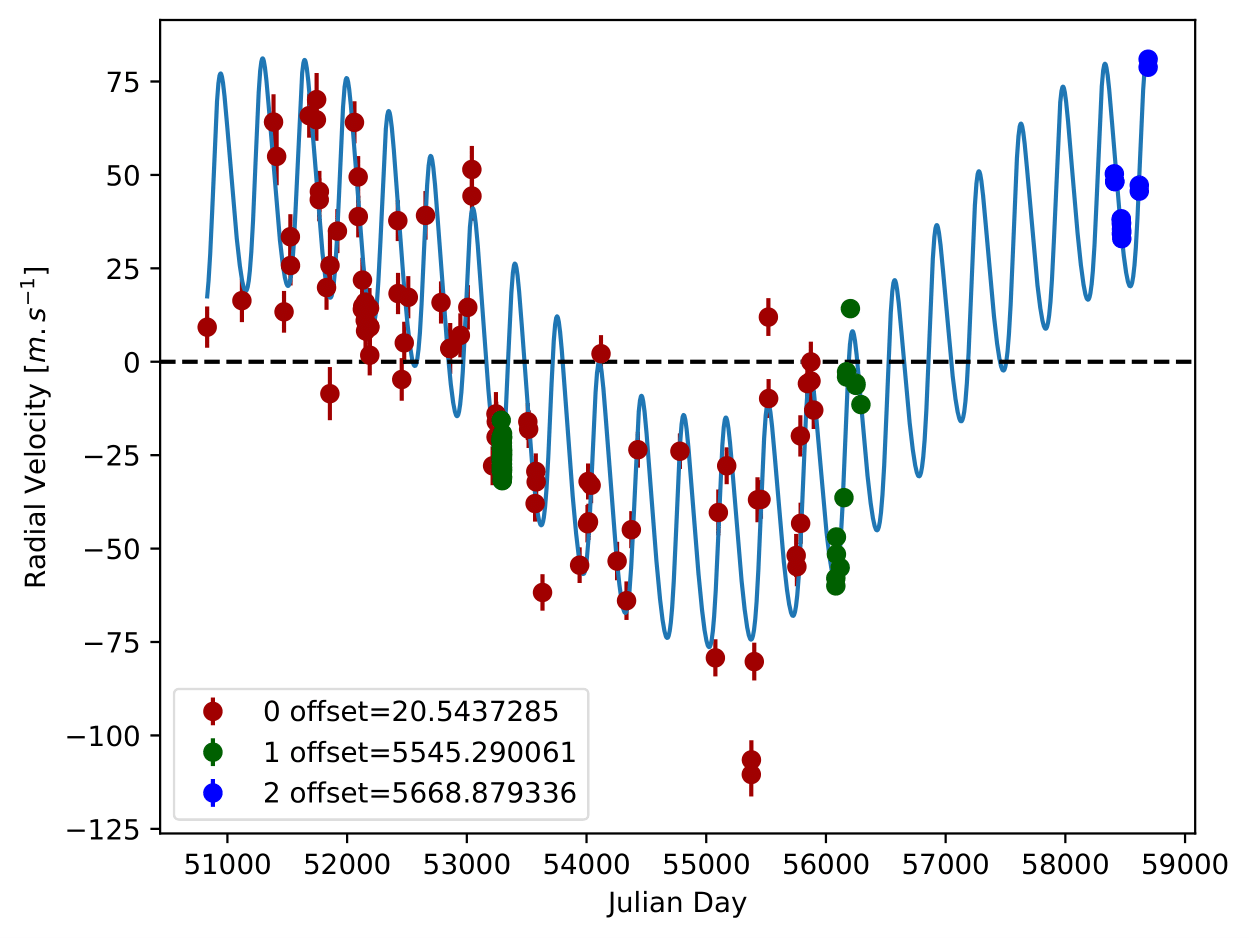}
\includegraphics[width=0.45\textwidth]{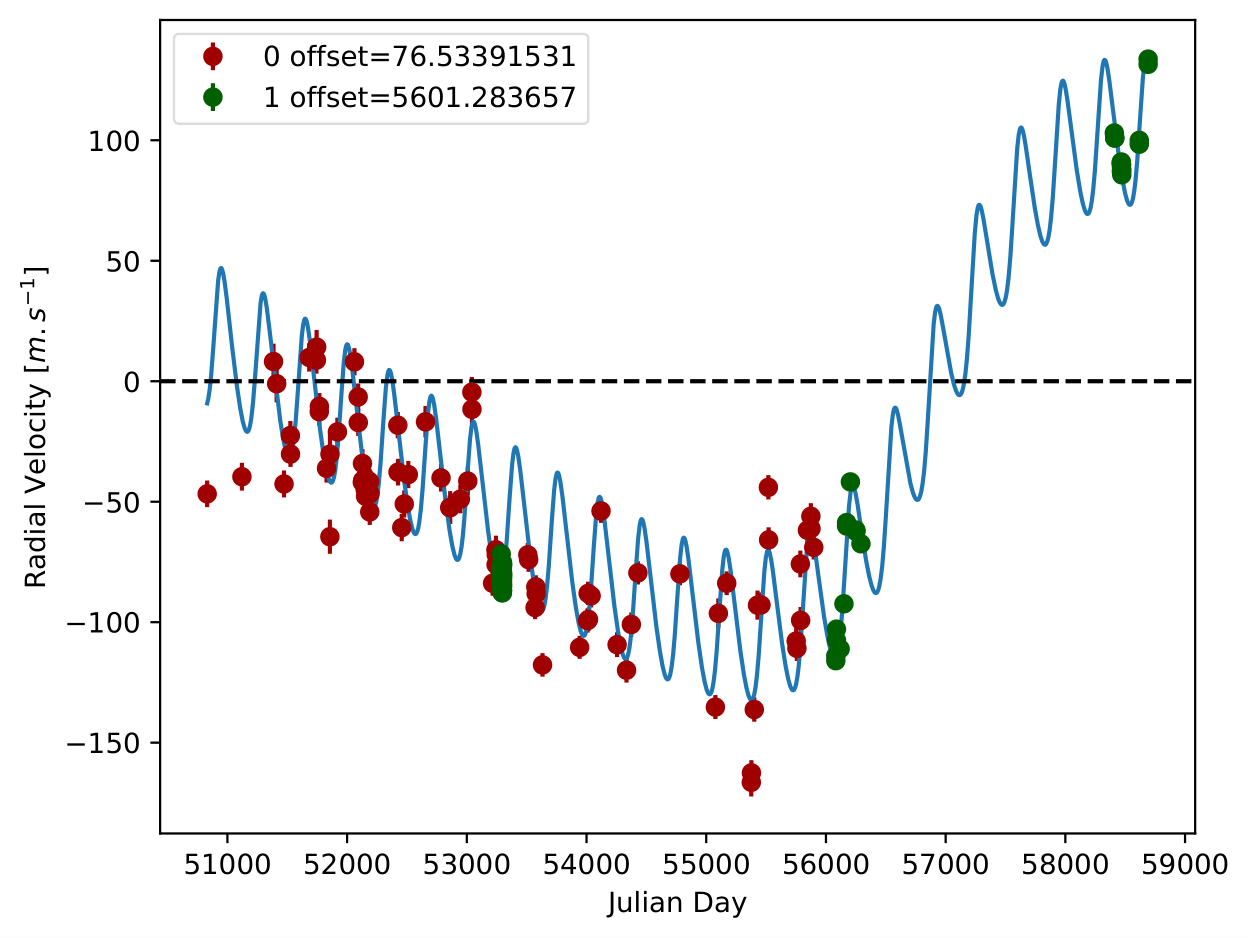}
\includegraphics[width=0.45\textwidth]{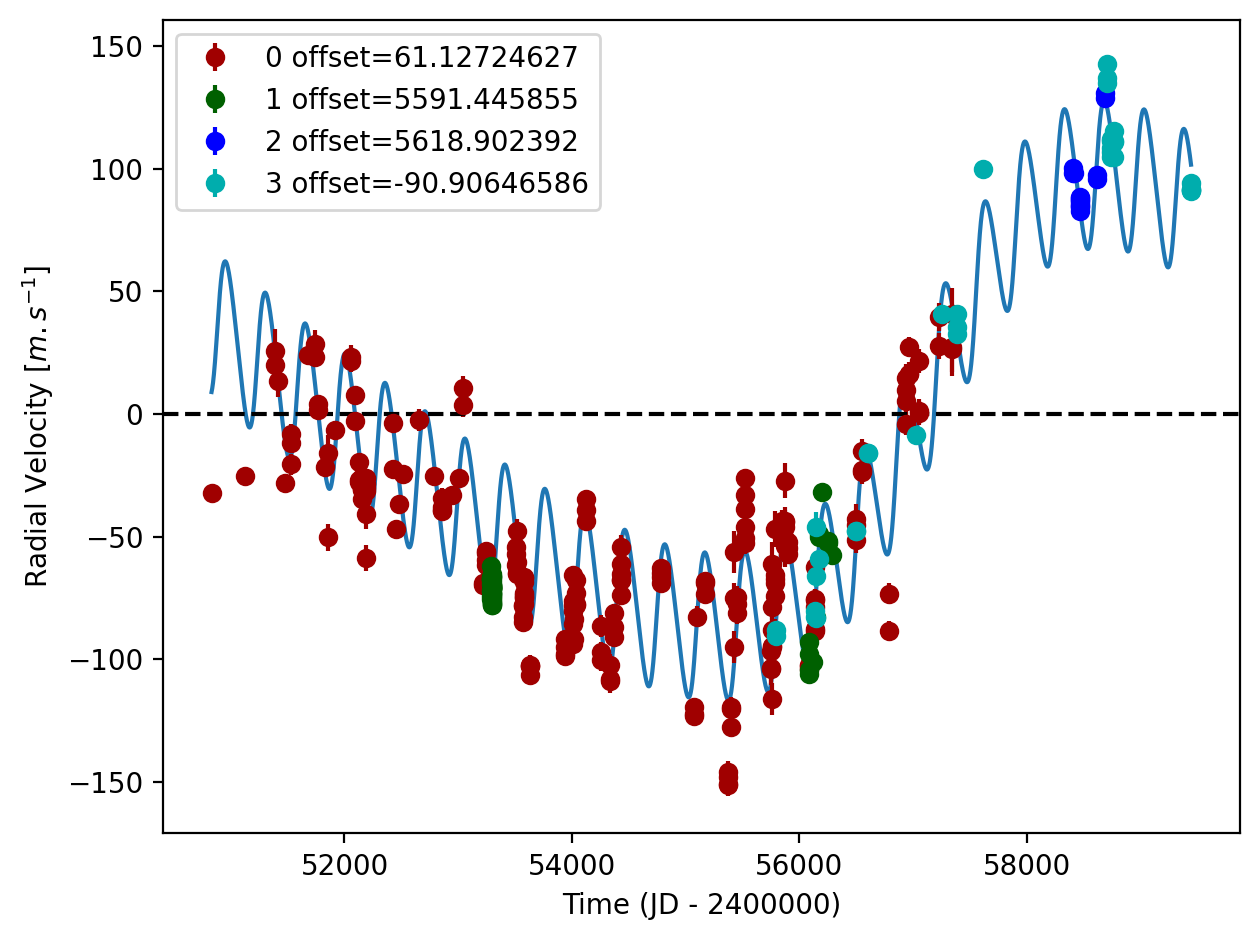}
\caption{Impact of the instrumental offsets. Case of HD 142A. \textit{Top}: fit (blue curve; genetic algorithm; see GitHub for the MCMC fits) of the RV with the instrumental offset between the two HARPS datasets taken as a free parameter. Red – AAT; green – HARPS pre-upgrade; blue – HARPS post-upgrade. \textit{Middle}: fit of the RV data assuming an instrumental offset of 14.9 m/s between the two HARPS datasets. Red – AAT; green – HARPS. \textit{Bottom}: fit of all the RV data available. Red – AAT; green – HARPS pre-upgrade; blue – HARPS post-upgrade; cyan - MIKE.
\label{ex_HD142}} 
\end{figure}

When the RV time series are obtained using different instruments, the fitting procedure requires considering one instrumental RV offset for each instrument, in addition to the usual set of parameters (orbital elements, planet mass, star RV, and jitter). This is also true when significant changes occur on a given instrument. This may lead to degenerate solutions if the data only covers part of the planet’s period or if there is no overlap between the data obtained with different instruments. 
An example is given in Figure \ref{ex_HD142}, for the case of HD 142A, which is observed with HARPS prior to and after a major instrumental upgrade in 2015 \citep{2015Msngr.162....9L} that induced non-negligible, spectral-type-dependent RV offsets. 
This instrumental offset cannot be constrained properly with the data available at the time of the CH survey, due to the large temporal gap between the two HARPS datasets.

Using these data, and considering the instrumental offset between the two HARPS data sets as a free parameter, we find a semi-major axis is 7.7 au, its minimum mass of 5.1 \Mjupv, an eccentricity close to 0; and an instrumental offset of 123.6 m/s. This value is considerably higher than that (14.9 m/s) measured for similar stars \citep{2015Msngr.162....9L}. Fixing instead an instrumental offset of 14.9 m/s between the two HARPS data sets, the semi-major axis becomes 10.3 au, the minimum mass 11.7 \Mjupv, and the eccentricity 0.34. Both solutions lead to a similar root mean square (hereafter rms) of the residuals (9.6 and 10.4 m/s). Note that new AAT and MIKE data have recently been published recently by \cite{2022ApJS..262...21F}, covering a similar period as the HARPS data. Using these data in addition to the previous ones, we find a semi-major axis of 9.6 au, a minimum mass of 10.4 \Mjupv, an eccentricity of 0.27, and an offset of 27 m/s. 
It is therefore critical that the temporal periods of the RV data obtained with different instruments overlap, at least partially, to properly constrain the different instrumental offsets.

\subsection{Fitting tools}

Finally, the detection and characterization of RV planets rely on approaches and tools that are not perfect. The orbital and mass characterization relies on genetic algorithms \footnote{Genetic algorithms fit RV data with one or several companions using an evolutionary algorithm where the genes are the Keplerian elements. They make use of a loose sub-population segregation for maintaining the gene pool diversity and thus promoting wider parameter exploration.} (CH survey), or on Bayesian MCMC statistical tools (CL survey) whose exploration may be limited, and sometimes biased \citep{2011MNRAS.410...94G}, especially in case of sparsely sampled data \citep{2014ApJ...781...35P}. For instance, in the case of HD 26161, solutions with very high eccentricities (up to 0.98) and \textit{a} (up to ~240 au) yield rms of residuals (genetic algorithms) or likelihoods (MCMC) comparable to a large range of solutions with lower eccentricity and semi-major axis values. Yet, such solutions are not captured by the genetic algorithm or the MCMC.

As an additional illustration of the limits of the genetic algorithm or the MCMC in case of (very) poor temporal coverage of the orbits, we compute the RV time series induced by a fake, high eccentricity (0.95), 70 \Mjup companion planet, corresponding to one plausible solution with \textit{a} = 240 au found when fitting the HD 26161 RV data. We use the calendar of observations of HD 26161. Noise is added in the following way: for each epoch, the synthetic RV is drawn from a normal distribution centered on the theoretical value of the RV and with a 1-$\sigma$ width corresponding to the measurement uncertainties. With priors on \textit{a} in the range 0-300 au, the MCMC posteriors are in the range $\sim$20-30 au, eccentricity 0.56-0.67, and minimum mass 40-110 \Mjupv. Hence, the MCMC fails to find the actual orbital parameters and minimum mass of the fake planet. 

\section{Revisiting the CH and CL surveys. Approach}

\subsection{Data}
As mentioned above, we focus on the LPGPs with \textit{a} greater than 3 au (CH survey) and \textit{a} greater than 5 au (CL survey). Among these 22 (resp. 19) planets, 41\% (resp. 47\%) have reported orbital periods (much) longer than the actual temporal baseline of the available data\footnote{Planets with \textit{a} < 3 (resp. 5 au) are in general monitored over their whole orbital period.}. Hence, the risk of a poor characterization of these long-period giant planets is high. We, therefore, revisit these targets, using i/ the RV data used in the surveys plus, when available, additional RV data, and ii/ fitting tools similar to those used in the published analysis, i.e. a genetic algorithm (CH survey), and an MCMC (CL survey).

\subsubsection{CH survey data}

We focus here on the 22 LPGPs with a semi-major axis greater than 3 au. We also consider HD 142A, for which the CH survey detected a short-period (350 d) planet, and for which an LPGP was detected after the publication of the CH survey. Each RV data set includes data from several (up to seven) instruments. As the data used by the CH survey were not published in that paper, we use data available in the literature, in the ESO archive, or in the DACE database \footnote{https://dace.unige.ch}. Additional data are obtained for 22 systems (only HD 142022 was not re-observed). The new data are included in our new analysis.

\subsubsection{CL survey data}

We focus here on the 19 LPGPs with a semi-major axis greater than 5 au. We use the data published in the CL survey. Whenever available, additional public data is also considered for the present analysis. However, in no cases do they increase the time baseline. Again, each RV data set includes data from several (up to seven) instruments. 

\subsection{Fitting procedures}

The data are first fitted using a genetic algorithm, DPASS \citep{2019NatAs...3.1135L}, as is done by the CH Survey. DPASS provides only the best solution, corresponding to the smallest rms of the RV residuals. MCMC fitting is also performed, as in the CL study. Unlike genetic algorithms, it estimates uncertainties associated with the various parameters of the solution. The MCMC sampling tool is based on the emcee 3.0 library \citep{2013PASP..125..306F}, using a mix of custom move functions to alleviate potential multi-modality problems and the cyclicity of angular variables. 

For both DPASS and MCMC, the free parameters considered are the orbital parameters (semi-major axis, eccentricity, argument of periastron and phase), the minimum mass(es) of the planet(s), the star RV (global offset due to the star systemic velocity), a stellar jitter to take into account the short-term (up to several days) stellar variations, and additional instrumental offsets. One instrumental offset is considered for each instrument, unless major changes occurred during their lifetimes. This is the case for HARPS, which had a fiber upgrade in 2015, HIRES, which had a detector change in 2004, CORALIE which had significant upgrades in 2007 \citep{2010A&A...511A..45S} and 2014 and SOPHIE which had significant upgrades in 2011. The instruments are therefore regarded as different instruments before and after these upgrades. HARPS before and after the upgrades will be referred to as H03 and H15, respectively. HIRES before and after the 2004 upgrades \citep{2019MNRAS.484L...8T} will be referred to as Hir94 and after Hir04, respectively. CORALIE will be referred to as C98 before the 2007 upgrade, C14 after 2014, and C07 in between, and SOPHIE before and after the 2011 upgrades \citep{2013A&A...549A..49B} will be referred to as SOPHIE and SOPHIE+, respectively.

The priors associated with the LPGPs are very loose so as to explore a wide range of possibilities. Uniform priors are considered for all fitting parameters except for targets with poorly constrained periods. In these cases, a log-uniform prior is considered for the semi-major axis in order to allow a better exploration of large \textit{a} solutions. In the case of systems hosting one or several well-characterized shorter-periods planets, in addition to the LPGP, the priors related to the short-period planets are chosen in the vicinity of the actual orbital parameters and minimum masses, to help obtain a faster convergence.

Depending on the temporal coverage of the RV variations, we use different strategies to characterize or constrain the orbital parameters of the companions, as described hereafter.

\subsubsection{Cases of good temporal coverage}

When the temporal coverage of the data is good, there is no degeneracy between orbital solutions and star RV; the MCMC provides good estimates of the orbital solutions.

An example is the case of HD 117207 which is a 1.04 \Msun, G8 IV star \citep{2005ApJ...619..570M}. Based on 42 RV HIRES measurements obtained between 1997 and 2004, \cite{2005ApJ...619..570M} reported a giant planet signal with a period of $2627 \pm 63.51$ days, a minimum mass of 2.06 \Mjup and an eccentricity of $0.16 \pm 0.05$. The CH survey reported properties for the HD 117207b close to those reported in \cite{2005ApJ...619..570M}.

\begin{figure}[t!]
 \centering
\includegraphics[width=0.45\textwidth]{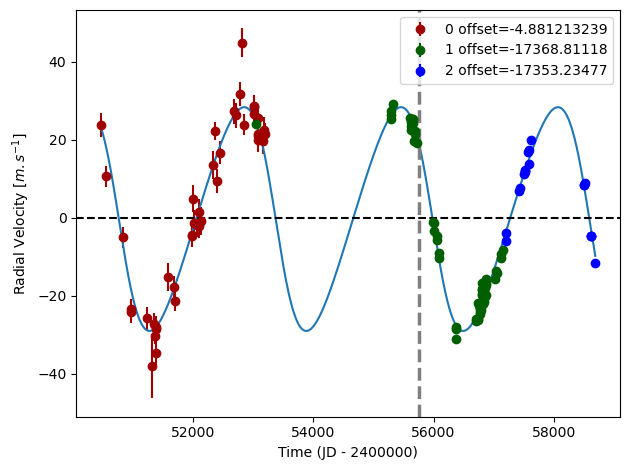}
\includegraphics[width=0.45\textwidth]{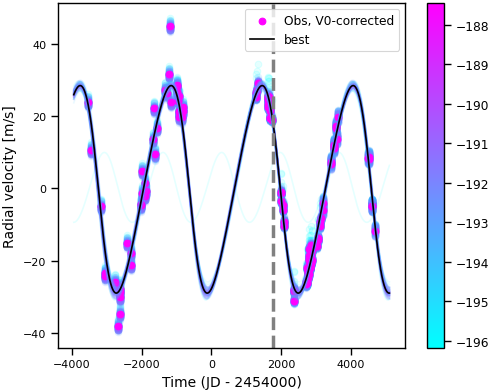}
\caption{Orbital fits for HD 117207 b. \textit{Top}: fit of the HD 117207 RV with DPASS. Red - Hir94; green - H03; blue - H15. The blue curve shows the best fit. \textit{bottom}: fit of the HD 117207 RV using MCMC. The black curve shows the best fit. The colorbar corresponds to the log-likelihood of the fits. The gray dotted line indicates the end of the CH survey.
\label{fit_HD117207}} 
\end{figure}

In the present study, in addition to the dataset of \cite{2005ApJ...619..570M}, 96 RV HARPS measurements obtained between 2004 and 2019 are used. DPASS and MCMC (1000 walkers and 300000 iterations) are used to fit the data. The fits are shown in Fig \ref{fit_HD117207}. The properties used in the CH survey for HD 117207b are confirmed.

\subsubsection{Cases of poor temporal coverage}

When the temporal coverage is poor, the solutions may be degenerate. This is, of course, not captured by the genetic algorithms, which only provide the best solution (corresponding to the lowest rms of the residuals). The MCMC offers a better exploration, but there is no guarantee that all the possible solutions are sampled in case of poor temporal coverage. Two important parameters identified (see Section 2) are the star RV offset and the semi-major axis of the planet. We explore their impact by fixing the RV offset to arbitrary but plausible values given the star RV uncertainties provided by Gaia or constraining the prior range of semi-major axes, and fitting the data with DPASS or MCMC. We compare the rms of the residuals (DPASS) or the likelihoods (MCMC) with the ones obtained with the RV offset or \textit{a} free. When comparable values are obtained, we conclude that the solutions are equally plausible. This leads to sometimes considerably wider ranges of possible solutions.

\begin{figure*}[t!]
 \centering
\includegraphics[width=0.32\textwidth]{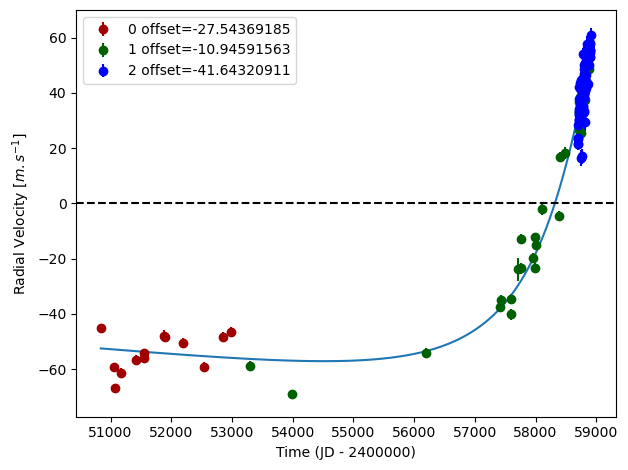}
\includegraphics[width=0.32\textwidth]{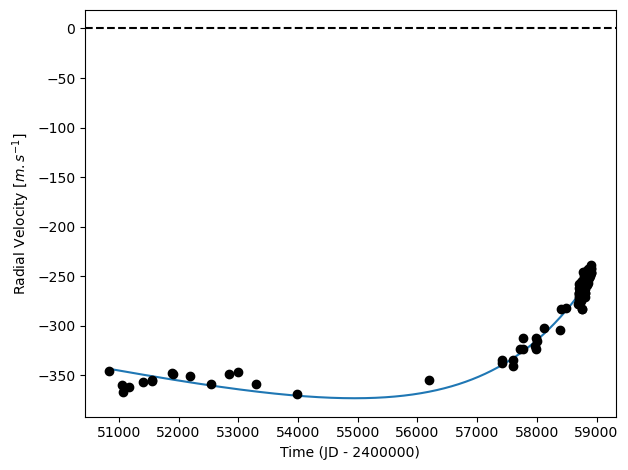}
\includegraphics[width=0.32\textwidth]{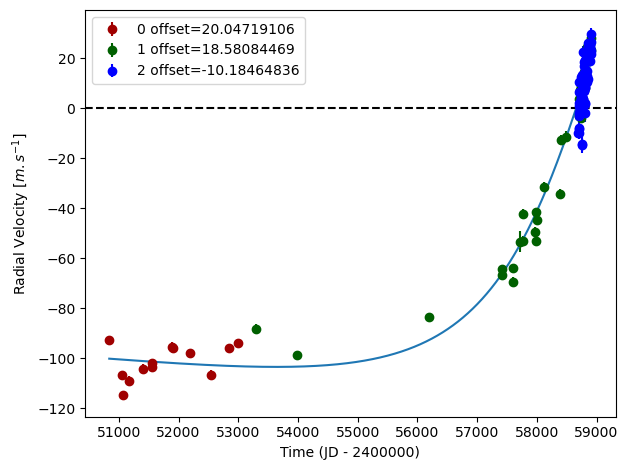}
\includegraphics[width=0.32\textwidth]{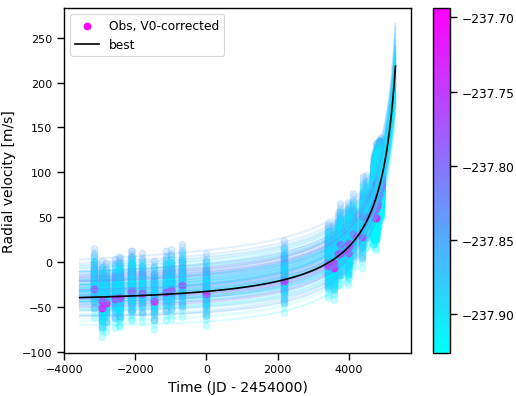}
\includegraphics[width=0.32\textwidth]{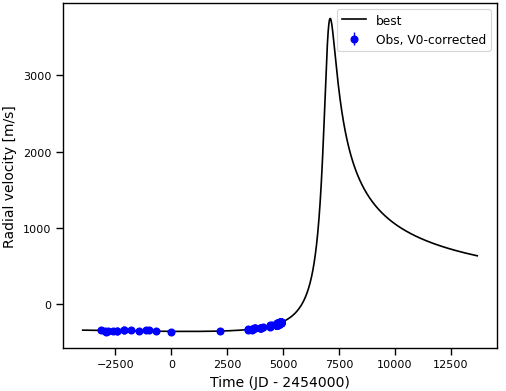}
\includegraphics[width=0.32\textwidth]{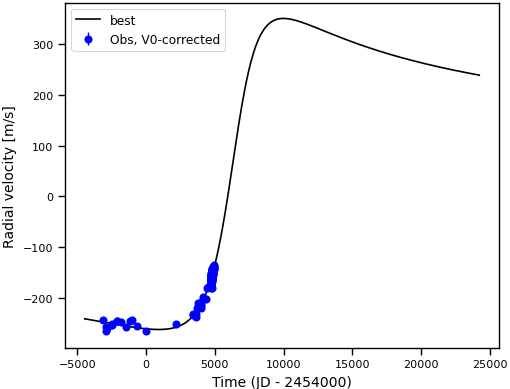}
\caption{Orbital fits for HD 26161b. \textit{Top left}: fit of the HD 26161 RV with DPASS. Red – Hir94; green – Hir04; blue – APF. The blue curve shows the best fit. \textit{Bottom left}: fit of the HD 26161 RV using MCMC. The black curve shows the best fit. The colorbar corresponds to the log-likelihood of the fits. \textit{Top middle}: fit of the HD 26161 RV with DPASS, with a subtracted stellar offset fixed to 300 m/s. Black points correspond to the data corrected from the instrumental offsets. The blue curve shows the best fit. \textit{Bottom Middle}: best fit of the HD 26161 RV using MCMC, with a subtracted stellar offset fixed to 300 m/s. The black curve shows the best fit. \textit{Top right}: fit of the HD 26161 RV with DPASS, with the minimum \textit{a} fixed at 240 au. The points are the same as on the \textit{Top left}. The blue curve shows the best fit. \textit{Bottom right}: best fit of the HD 26161 RV with MCMC, with a prior on \textit{a} in the range 240-300 au. The black curve shows the best fit.
\label{fit_HD 26161}} 
\end{figure*}

An example is provided in the case of HD 26161 which is a 1.13 \Msun, G0 star \citep{2021ApJS..255....8R}. Using 50 RV HIRES measurements obtained between 1998 and 2019 and 84 RV APF measurements obtained between 2019 and 2020, the CL survey reported an LPGP with a period of $32000_{-10000}^{+21000}$ days, a minimum mass of $13.5_{-3.7}^{+8.5}$ \Mjup and an eccentricity of $0.82_{-0.05}^{+0.061}$. We use DPASS and MCMC (1000 walkers, 400000 iterations, \textit{a} between 0 and 1000 au, RV offset between -1 and 1 km/s, eccentricity between 0 and 0.99) to fit the data. DPASS finds a companion with a period of 64224 days, a minimum mass of 14.1 \Mjup, and an eccentricity of 0.78, with a corresponding rms of residuals of 6.3 m/s. With the MCMC, solutions with a period peaking at about 100 yr and extending to more than 300 yr (peak of \textit{a} at about 14 au, extending to more than 30 au), a minimum mass between 13 and 170 \Mjup, and an eccentricity larger than 0.75 are found. The fits are shown in Figure \ref{fit_HD 26161}.

Yet, as no extremum is covered with the present dataset, the stellar offset, and therefore the companion orbital properties, are not well constrained. To explore the impact of the RV offset on the results of the genetic algorithm, we fix the RV offset to arbitrary values up to 300 m/s (compatible with the range of published values) and fit the data with DPASS. Very different solutions are found, with rms of residuals very close to that obtained with a free offset. For instance, with an RV offset of 300 m/s, we find a semi-major axis of 42 au, a minimum mass of 80 \Mjup, and an eccentricity of 0.66, associated with an rms of 7.2 m/s, comparable to the rms obtained with a free RV offset. 

A similar exercise is done by fixing various semi-major axes and similar conclusions are reached. It appears that with \textit{a} up to 240 au, solutions giving rms of residuals within 1 m/s from the baseline (7 m/s) are obtained. An example is shown in Fig. \ref{fit_HD 26161}.

We then check the impact of the RV offset or the semi-major axis on the MCMC results. We perform a new MCMC sampling fixing the RV offset at 300m/s. The prior on \textit{a} is, again, very loose, 0 to 1000 au. The \textit{a} distribution peaks at about 40 au, significantly higher than in the case where the RV offset is free. The best likelihood is, however, only slightly lower than that obtained with a loose prior on the RV offset, and the distributions of the likelihoods overlap. 

Finally, we run an MCMC with the prior on the \textit{a} constrained to 240-300 au. As expected, the distribution of a sample by the MCMC is truncated at the min value, 240 au, but the distribution of likelihoods overlaps with the one obtained with very loose priors for the \textit{a} in the range 0-1000 au. Hence, some solutions sampled with \textit{a} in the range 240-300 au provide likelihoods identical to those sampled with loose priors on \textit{a}. These solutions are equally likely, but the MCMC does not capture them when the priors on \textit{a} are very loose.

This analysis shows that neither the DPASS nor the MCMC captures all the solutions, especially those with high eccentricities or high semi-major axis\footnote{Algorithms such as the JOKER \citep{2017ApJ...837...20P} used in the CLS do not fully alleviate this bias either.}. 

In most cases of poorly covered RV, however, the RV time series includes one (putative) extremum. The solutions are less degenerate than in the case of HD 26161, depending on the number of instruments used, and the level of overlap between the instruments. An illustration is given in the case of HD 66428 (1.05 \Msun, G5 star; \cite{2021ApJS..255....8R}), a system with an inner, well constrained, planet (HD 66428b) and an outer companion (HD 66428c), whose RVs are poorly constrained. Using 99 RV HIRES measurements obtained between 2000 and 2019, The CL survey reported for HD 66428b a period of $2288.9_{-6.8}^{+6.1}$ days, a minimum mass of $3.19 \pm 0.11$ \Mjup, and an eccentricity of $0.418_{-0.014}^{+0.015}$ and, for HD 66428c, a period of $39000_{-18000}^{+56000}$ days, a minimum mass of $27_{-17}^{+22}$ \Mjup, and an eccentricity of $0.32_{-0.16}^{+0.23}$. 

In the present study, we use in addition to the CL survey's data set, 28 RV HARPS measurements obtained between 2004 and 2015. DPASS and MCMC (1000 walkers and 400000 iterations) are used to fit the data. The properties found for HD 66428b are similar to those reported in the CL survey. For HD 66428c, a period of 9107 days, a minimum mass of 2 \Mjup, and an eccentricity of 0.19 are found with DPASS, with a corresponding rms of residuals of 2.8 m/s, and a period between 27000 and 125000 days, a minimum mass between 22 and 296 \Mjup and an eccentricity between 0.38 and 0.78 are found using MCMC. As the RV curve of HD 66428c only covers a minimum, the period (or \textit{a}) is actually not well constrained. 

\begin{figure*}[t!]
 \centering
\includegraphics[width=0.45\textwidth]{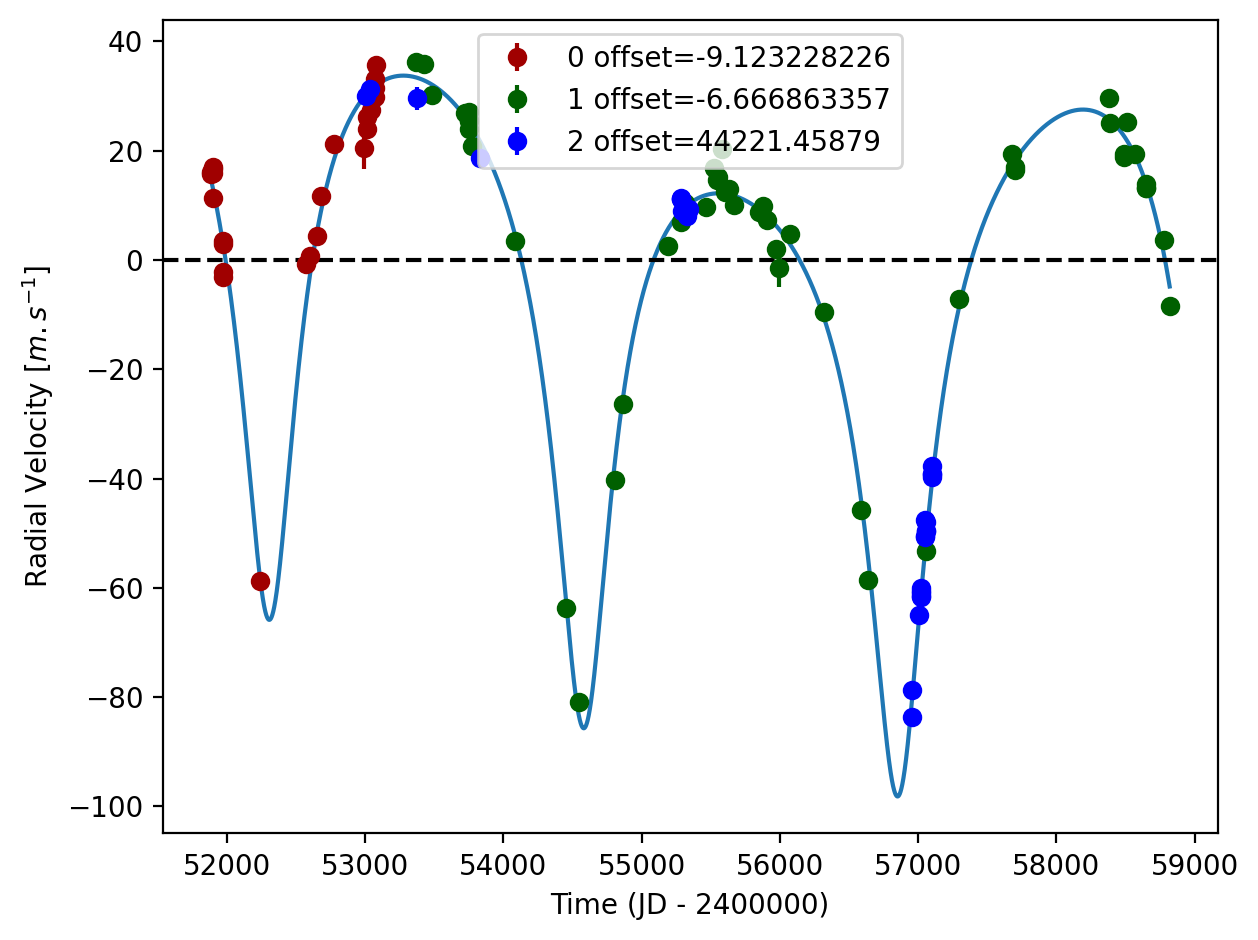}
\includegraphics[width=0.45\textwidth]{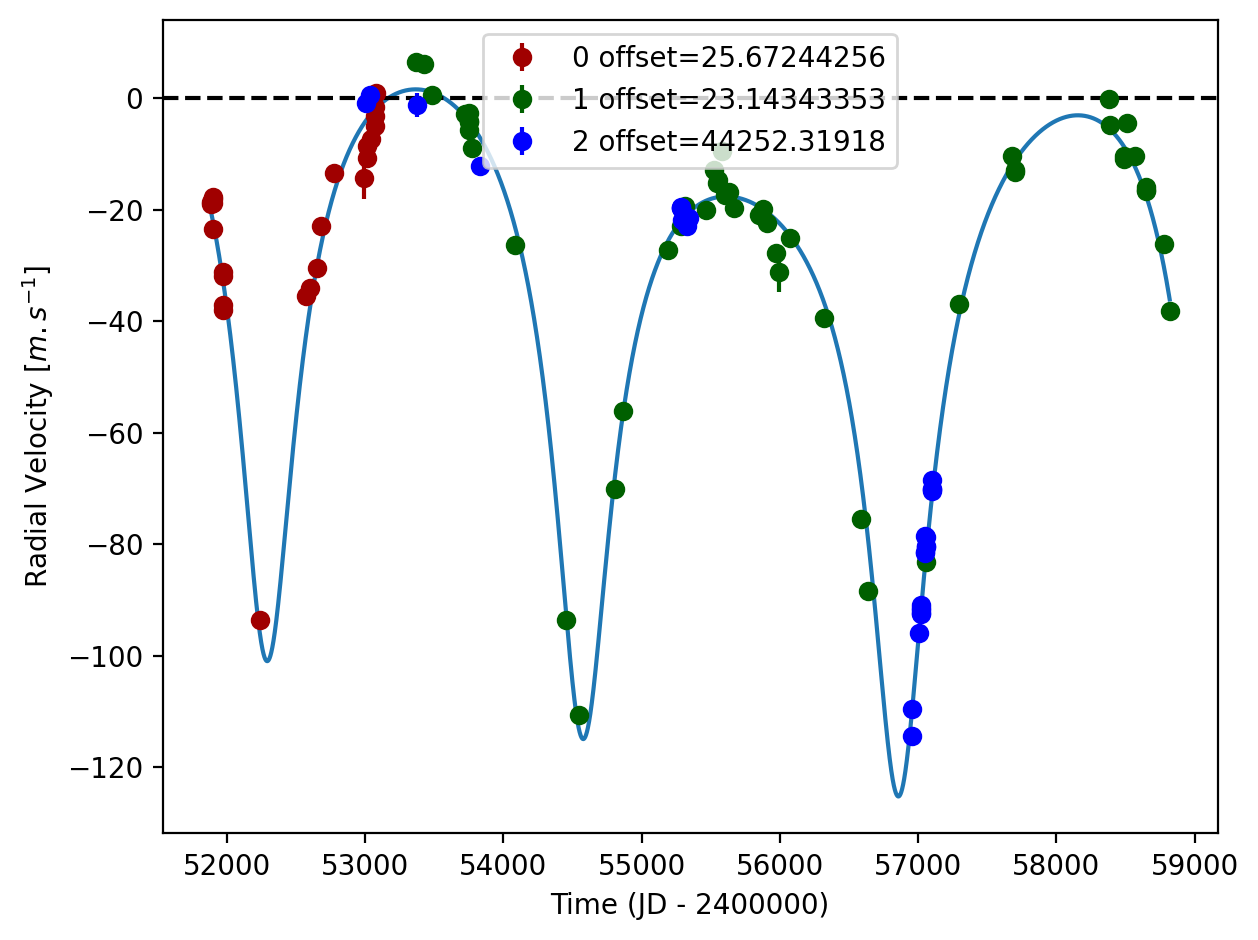}
\includegraphics[width=0.45\textwidth]{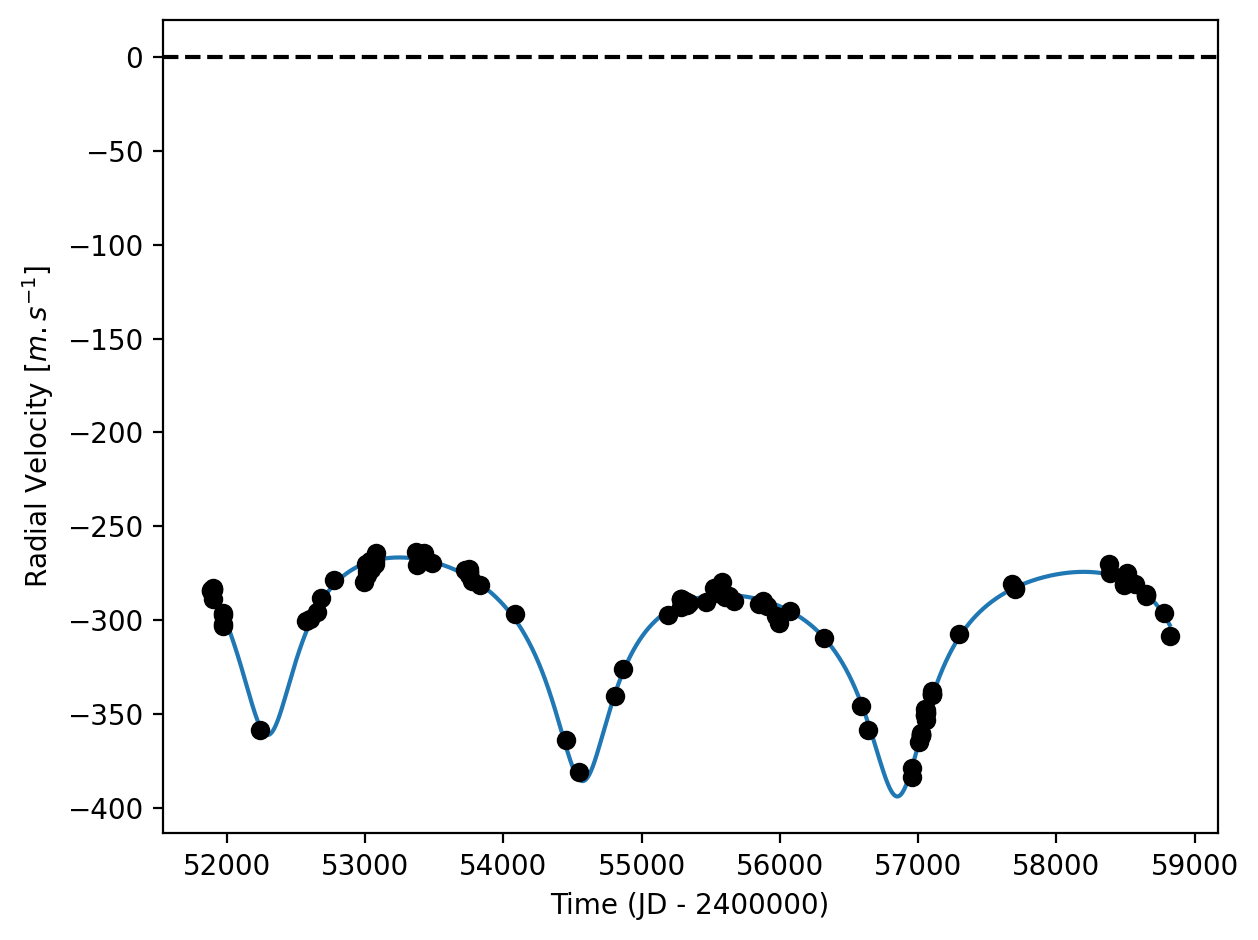}
\includegraphics[width=0.45\textwidth]{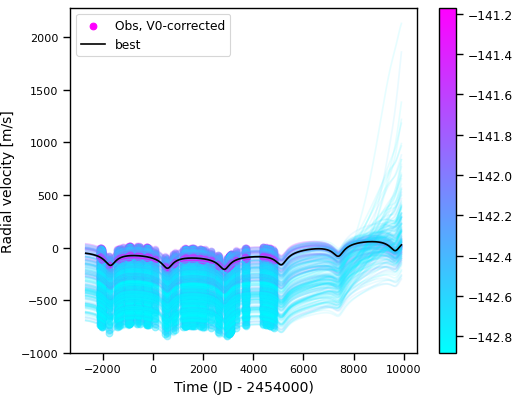}
\caption{Orbital fits for HD 66428 b and c.\textit{Top left}: fit of the HD 66428 RV with DPASS. Red – Hir94; green – Hir04; blue – H03. The blue curve shows the best fit. \textit{Top right}: fit of the HD 66428 RV with DPASS, with the minimum \textit{a} fixed at 190 au for HD 66428c. The points are the same as on the left. The blue curve shows the best fit. \textit{Bottom left}: fit of the HD 66428 RV with DPASS, with a stellar RV offset subtracted fixed to 300 m/s. Black points correspond to the data corrected for the instrumental offsets. The blue curve shows the best fit. \textit{Bottom right}: fit of the HD 66428 RV using MCMC. The black curve shows the best fit. The colorbar corresponds to the log-likelihood of the fits.
\label{fit_HD66428}} 
\end{figure*}

To explore the range of possible values, we fix the semi-major axis to different values and fit the data with DPASS. Semi-major axis up to 190 au does not significantly change the rms of the residuals (3.3 m/s against 2.8 m/s with a left free). In this case, the minimum mass is 4 \Mjup and the high eccentricity is 0.95, corresponding to the maximum value of the eccentricity prior.
To test the impact of the stellar RV offset, we also fix this offset to different values and fit the data, once corrected from the instrumental offsets for clarity purposes, with DPASS. It appears that stellar RV offsets up to 300 m/s do not significantly change the rms of the residuals (3 m/s against 2.8 m/s with stellar RV offset free). In this case, \textit{a} is 48 au, the minimum mass is 52 \Mjup and the eccentricity is 0.63. The fits are shown in Fig. \ref{fit_HD66428}.

\begin{figure*}[t!]
 \centering
\includegraphics[width=1\textwidth]{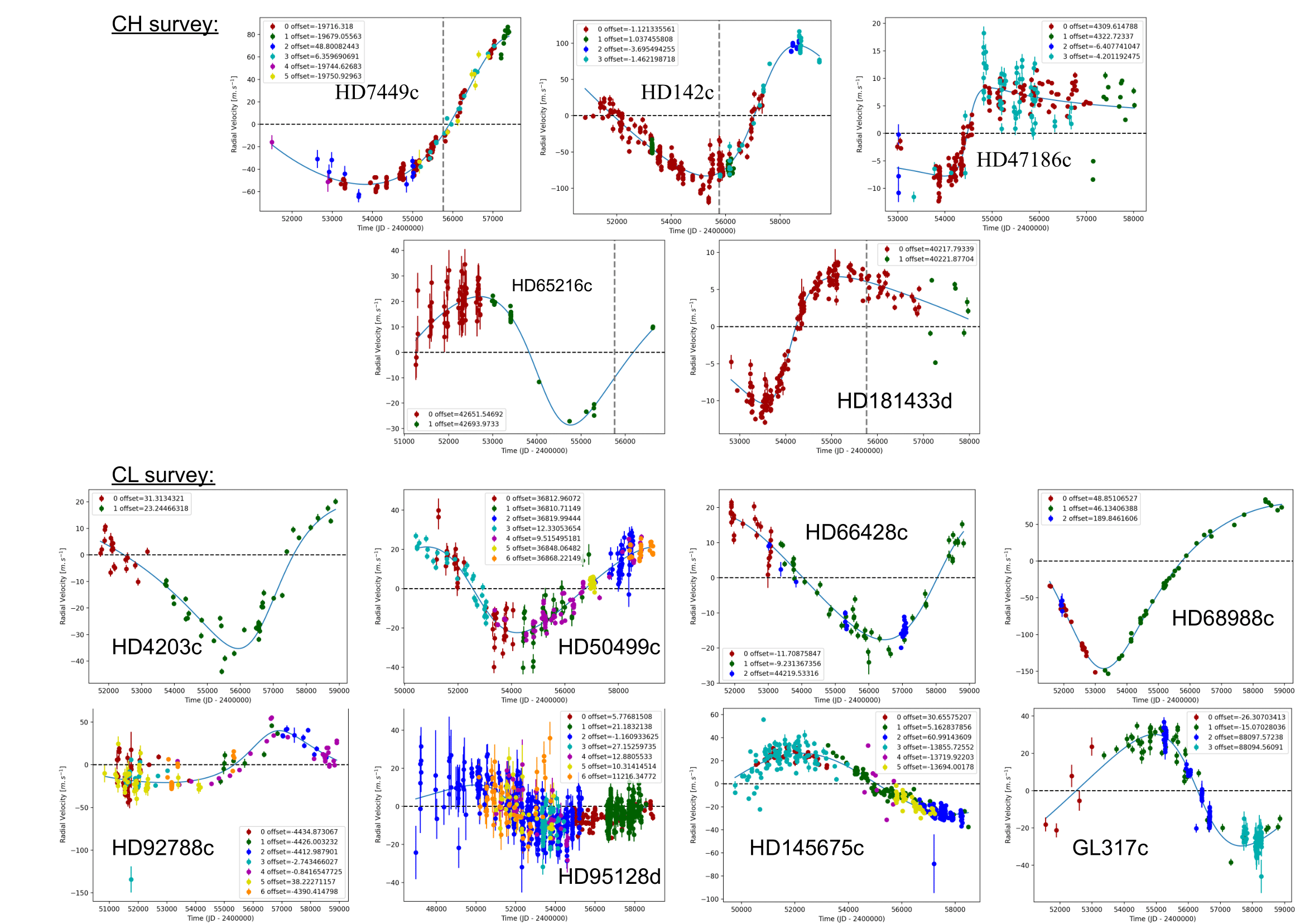}
\caption{Multiple systems with poorly constrained outer planets. RV residuals once the signal of the inner planet(s) has been removed. \textit{Top}: CH survey. The vertical gray dotted line indicates the end of the CH survey. \textit{Bottom}: CL survey. 
\label{multiple_system}} 
\end{figure*}

\subsubsection{Note on multiple systems}

The samples include several multiple systems: 13 in the CH survey and 13 in the CL survey (23 total systems). Once corrected from the signal(s) of the inner planet(s), it appears that in nine cases, the RV time-series cover (very) poorly the orbital periods of the LPGPs (see Figure \ref{multiple_system}), with none (one case, possibly two) or only one extremum (eight cases). In four other cases, the periods are roughly covered, but poorly sampled, and in two targets in the CH survey, the LPGPs are not detected. For these nine systems, we adopt a similar approach to that adopted in Section 3.2.2. When exploring the different orbital/mass solutions for the outer planet, the priors on the orbital parameters and the minimum mass(es) of the inner planet(s), which are well constrained by the RVs, are chosen close to the values obtained using loose priors.

\subsubsection{Note on stellar activity}
The stellar activity is not corrected, as in the CL study. This is a difference from the CH survey which corrected the RV from the stellar activity and will be discussed in the next section. 

\subsubsection{Note on Hipparcos-Gaia astrometry}

Among the 39 analyzed systems, 13 have actually been studied by coupling radial velocity data with absolute astrometry (\cite{2020A&A...642A..31D}, \cite{2021AJ....162..266L}, \cite{2022ApJS..262...21F}, \cite{2023A&A...670A..65P}). We do not use these data, however, because the LPGPs radial distributions that we discuss here are based on RV data only. Considering true masses in some cases only instead of \msini masses would break the homogeneity of the survey. 

\section{Revisiting the CH and CL surveys: Results}
We apply the procedure described in the previous section to each of our sample targets. A detailed analysis for each target is available on GitHub\footnote{$https://github.com/fphilipot/LPGP_supplementary$}. Illustrative examples of such analysis are provided in Appendix B. Figure \ref{LPGP_results} shows the revised periods and minimum masses, as well as the reported ones, for comparison purposes. The orbital parameters and masses found in the present study, together with those reported by the CH and CL analysis, are provided in Appendix D. 

\begin{figure*}[t!]
 \centering
\includegraphics[width=0.9\textwidth]{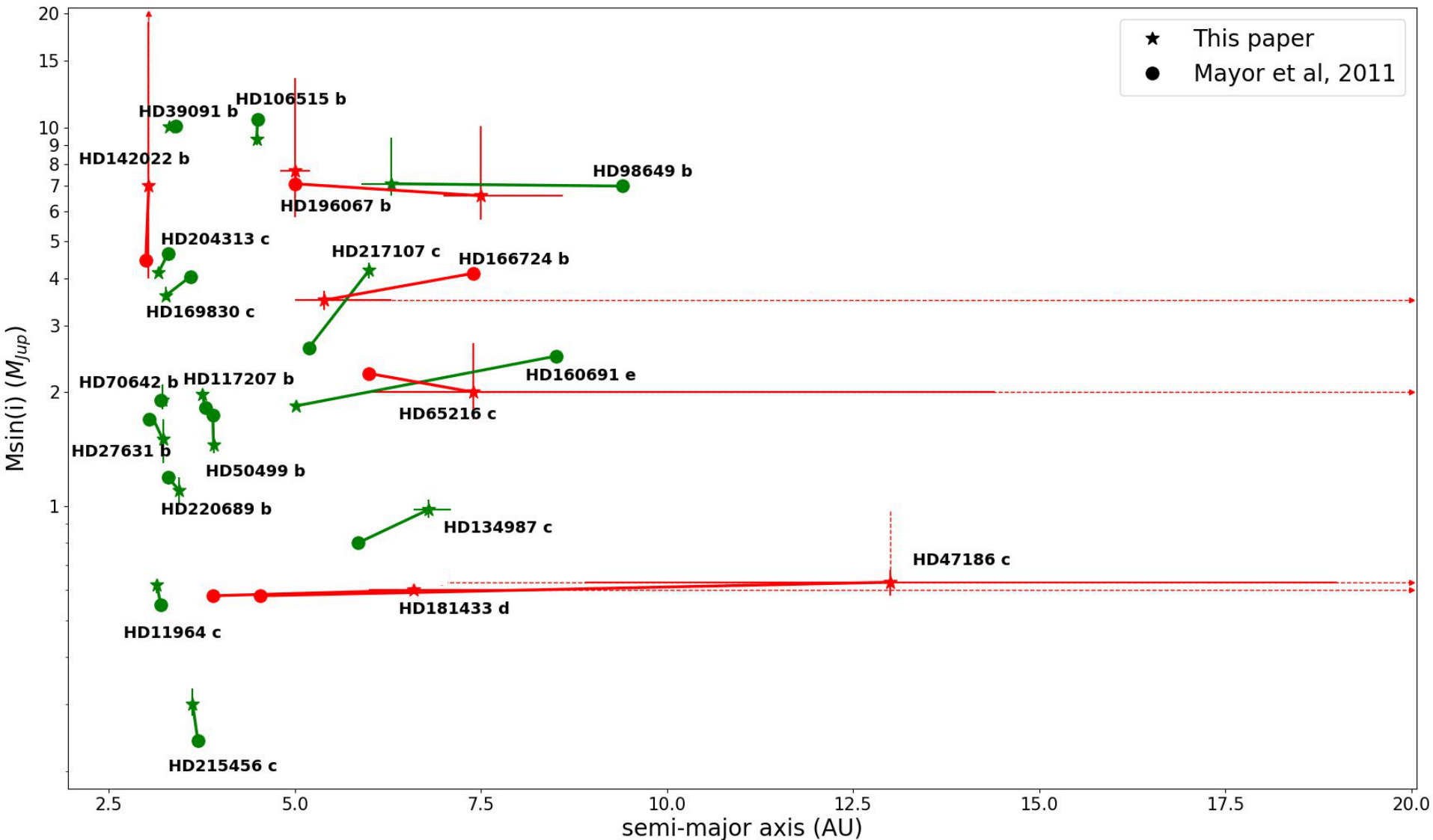}
\includegraphics[width=0.9\textwidth]{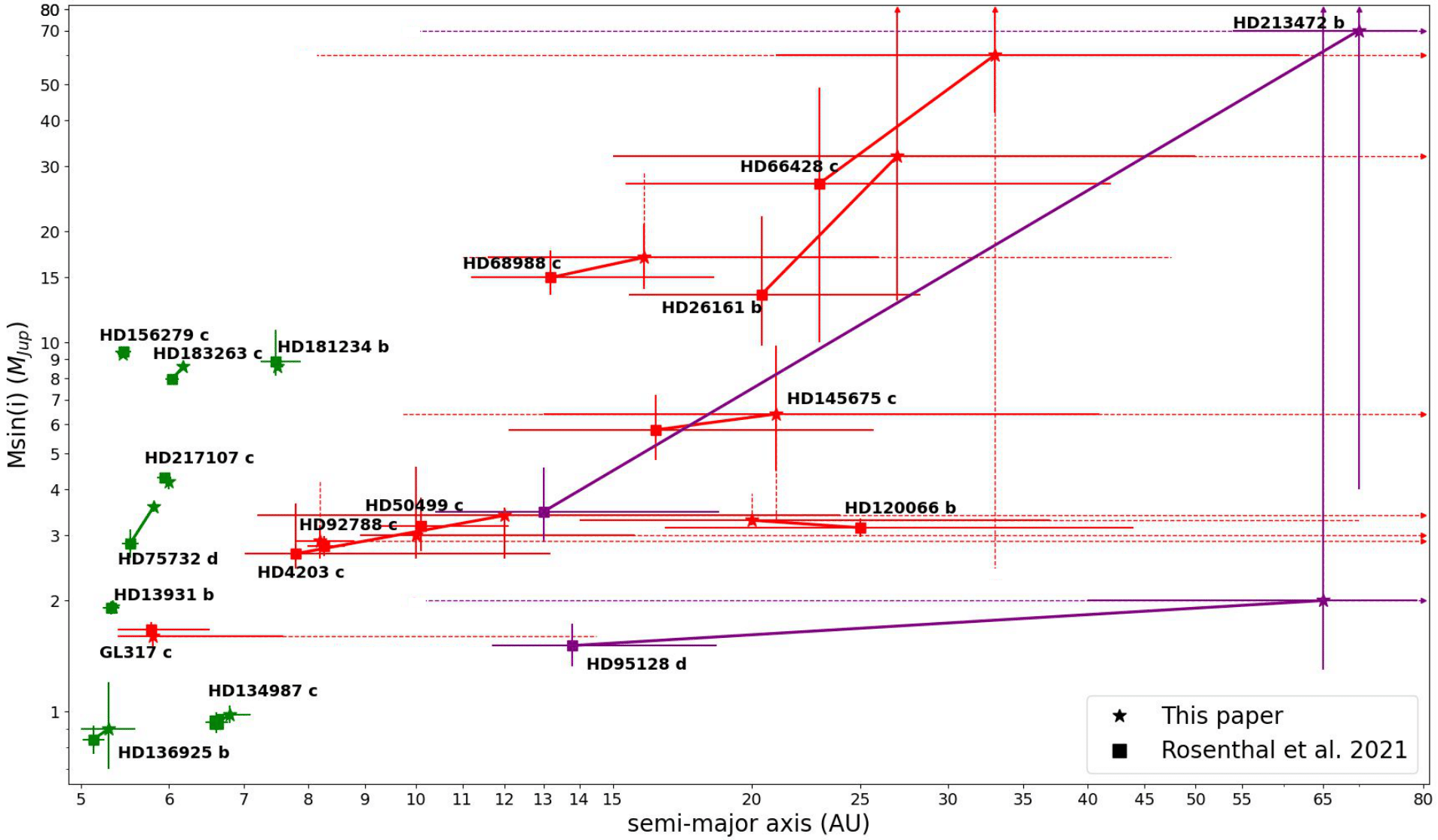}
\caption{Revised semi-major axes and minimum masses. \textit{Top}: Targets with reported \textit{a} > 3 au in the CH survey \citep{2011arXiv1109.2497M}. Lines between the revised (indicated by stars) and published values (indicated by filled circles) are drawn to facilitate the comparison between these values. Green colors correspond to planets for which the final orbital parameters and minimum masses are well constrained; red colors correspond to planets for which the semi-major axes or/and the minimum masses are not well constrained. The dotted lines correspond to solutions that provide rms of residuals within 1 m/s from the minimum rms. An arrow at the end of a dashed line means that values even beyond the limit can meet this criterion. HD 7449c is not shown as it is identified as a star. \textit{Bottom}: Same for the CL survey \citep{2021ApJS..255....8R} targets with reported \textit{a} > 5 au. The purple color corresponds to planets for which the semi-major axis and the minimum mass cannot be constrained.
\label{LPGP_results}} 
\end{figure*}

\subsection{CH planets (22) with \textit{a} > 3 au}

For 11 out of the 22 planets with a > 3 au, the orbital parameters and minimum masses estimated by the CH survey are confirmed. For the eleven others, the orbital parameters and minimum masses are significantly different from those reported (four cases), or a second solution is possible in addition to the one reported (one case), or the solutions are very degenerate and only very wide ranges of orbital parameters can be given (six cases). Note that for two stars, HD 142 \citep{2012ApJ...753..169W} and HD 50499 \citep{2019A&A...625A..71R}, two additional long-period planets have been reported in the literature, and another LPGP is suspected around HD 204313 \citep{2012ApJ...754...50R}. Finally, none of the eight planets with reported \textit{a} > 5 au has its orbital parameters/mass confirmed. We provide in Appendix B a detailed analysis of two targets for which we get significantly different from the CH analysis. 

We note that, as mentioned above, the present analysis does not correct the RV from stellar activity, while the results provided by the CH analysis do. 
However, as seen above, stellar activity has a significant impact on planets lighter than 0.6 \Mjup when considering planets orbiting beyond 3 au. For 18 stars out of 22, the masses reported by the CH analysis are greater than 0.6 \Mjup, hence, no strong effect due to stellar activity is expected. For two out of the four stars with reported masses smaller than 0.6 \Mjup, the present analysis provides results roughly similar to those reported, showing that activity does not play a strong impact on these targets. For the remaining two, significantly different results are found, mainly because of a significantly improved temporal coverage of the planet's orbits.

\subsection{CL planets (19) with \textit{a} > 5 au}

The results of the fits are in strong agreement with the published ones for the eight targets for which the RV monitoring is comparable or longer than the reported periods. seven out of these eight LPGPs have \textit{a} in the 5-7 au range and one has \textit{a} > 7 au\footnote{The only target with an initial \textit{a} < 7 au whose orbital properties are not confirmed is GL 317c. This is due to too sparse sampling of the variations in the first 2000 d.}. For the remaining 11 LPGPs, the present analysis finds, in general, (much) wider ranges of orbital parameters than those found by the CL survey (see example in Section 3.2.2). This is particularly true for the eight LPGPs with \textit{a} > 10 au. In fact, the MCMC (or genetic algorithms) happen to be heavily biased towards low semi-major axes for these poorly covered orbits. It is furthermore worth noting that five of the latter have a minimum mass possibly greater than 6000 $M_{Earth}$ (20 \Mjupv) and would therefore not be planets.

\section{Radial distribution of long-period giant planets beyond 3-5 au}

The CH and CL studies derived occurrence rates in bins of semi-major axes up to 10 au and 30 au respectively, for planets with masses between 30 and 6000 $M_{Earth}$ using i/ the actual detections, ii/ the estimated completeness of the surveys, and iii/ assumptions on the mass distribution of the planets. The radial distributions that fit the occurrence rates are then derived, using specific models. Important issues that impact the robustness of the results are discussed hereafter.

\subsection{Actual detections and characterizations}

As seen above, the current analysis of the CH and CL surveys data leads to significant revisions of the LPGPs orbital elements and minimum masses for planets with \textit{a} > $\sim $4-5 and 7 au, respectively. The possible companion masses are sometimes well above 6000 $M_{Earth}$, and the possible periods/\textit{a} go well beyond the last bins considered in the surveys (3594-10000 d for the CH analysis, and 16-30 au for the CL analysis). Furthermore, as discussed above, the ability to detect and correctly characterize low mass (typ. < 0.5 \Mjupv) giants orbiting beyond typically 3 au is, potentially, impacted by stellar activity.

\subsection{Survey completeness}

The CH analysis extrapolated the completeness derived in the CH survey \citep{2019ApJ...874...81F} to periods up to about 10000 d where the authors of the survey warned against deriving occurrences (and hence completeness) for periods longer than 10 yr (included in the last bin considered in the analysis). Hence, the completeness beyond $\sim$5 au is not robust. One should also note that the completeness provided by \cite{2019ApJ...874...81F} 
 was computed assuming circular orbits only.

In the CL analysis, the completeness was computed using injection-recovery of the signals from planets with different orbital parameters and masses in the RV time-series of each of the sample stars \citep{2021ApJS..255....8R}. To our understanding, “recovered” meant that the fitted period and semi-amplitude are within 25\%, and the phase within $\pi$/6, of the injected values. Their analysis was however affected by several biases:

\begin{itemize}

 \item The injected planets were drawn assuming a beta distribution of eccentricity strongly biased towards low eccentricities. This distribution was derived from the analysis of $\sim$400 RV planets detected in an earlier survey \citep{2013MNRAS.434L..51K}. However, in that survey, most of these planets had \textit{a} < 3-4 au; moreover, this beta distribution is only followed for the closest (\textit{a} < 1 au) planets, the eccentricities of the longer period planets following a flatter distribution instead. Hence, the assumption of this beta distribution for planets beyond 1 au is not justified. This leads to an overestimation of the recovery rates for planets with periods typically larger than the baseline of the survey ($\sim$21 years; median value), thus with \textit{a} > $\sim$6-8 au, compared to a uniform distribution. Indeed, assuming a uniform distribution for the eccentricities instead of the beta distribution would lead to considering a larger number of eccentric planets. High eccentricity planets with \textit{a} > 6-8 au produce RV that are $\sim$flat for long periods of time, in contrast with low eccentricity planets. Given the incomplete RV time coverage of LPGP, their detection probability is lower.
 \item No criterion relative to the fitted eccentricities was applied (i.e. no comparison between the fitted and injected eccentricities). Yet, using the tools used by the CL analysis to perform injection-recovery tests\footnote{available on github.com/California-Planet-Search/rvsearch}, it appears that requiring the absolute difference between the recovered and injected eccentricities to be less than 0.2 would decrease by 12\% the recovery rate of planets with \textit{a} > 8 au and by 20\% the recovery rate of planets with \textit{a} > 16 au. By applying a stricter criterion (e.g. an absolute difference less than 0.1), the recovery rate would decrease by 27\% for LPGPs with \textit{a} > 8 au and by 40\% for LPGPs with \textit{a} > 16 au.
 \item As discussed above, the MCMC fails to capture all the degeneracies between the RV of the star and the planet properties or to provide accurate orbital/mass solutions when the orbits are poorly covered. Given the median monitoring time of the survey (21 years), there is a potential impact on the results for planets with \textit{a} greater than typically 6-8 au.
 \item Finally, if the planets producing peaks close to the activity peaks were not considered as recovered because of possible confusion with stellar activity, one might expect a lower completeness, associated with larger uncertainties for planets orbiting in the 3-5 au range.

 \end{itemize}

 \subsection{Mass distribution}

 Both the CH and CL analysis assumes an identical mass distribution over the 30-6000 $M_{Earth}$ range for all semi-major axes. Such an assumption is not justified. 
 Indeed, the mass distribution for the 30-6000 $M_{Earth}$ range cannot be properly measured beyond 3-5 au in the CH survey and 8 au in the CL survey (see their figure 5), as low mass (typ. sub-Jupiter) planets cannot be detected at such distances. Furthermore, as seen above, detecting and characterizing planets with masses of less than about 0.3-0.6 \Mjup orbiting in the 3-5 au range is complicated due to stellar activity. Hence, occurrences for the whole 30-6000 $M_{Earth}$ range can only be measured robustly for semi-major axes below these values.

 \subsection{Impact on the radial distribution}

 At this stage, it is not possible to quantify the overall impact of these limitations and biases because of multiple unknowns/limitations (e.g. stellar activity, MCMC biases) and because the data that would be needed to do so are simply not available (e.g. the data needed to estimate the radial distribution of remote sub-Jupiter planets). 
 However, the present analysis shows that, despite the remarkable effort made in these studies, the radial distributions are not robustly assessed beyond $\sim$4 au for the CH survey and beyond 5-8 au for the CL survey. The 3-5 au range might even be poorly constrained given the possible impact of stellar activity, and above all, the poorly constrained mass distribution function for low-mass planets in this region.

\begin{figure}[t!]
 \centering
\includegraphics[width=0.49\textwidth]{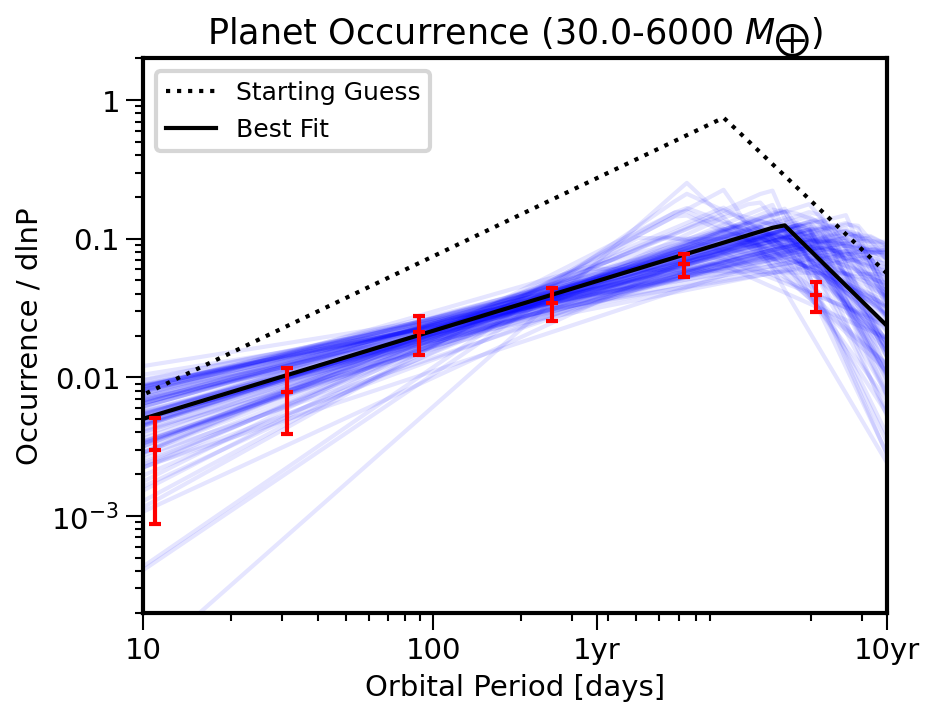}
\includegraphics[width=0.49\textwidth]{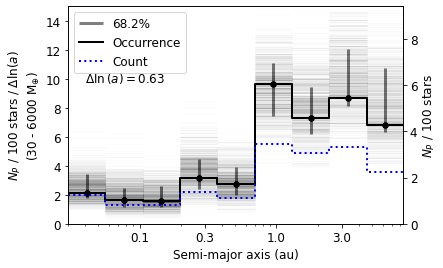}
\includegraphics[width=0.49\textwidth]{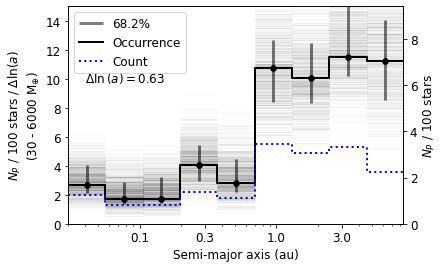}
\caption{Revised occurrence rates of giant planets. \textit{Top}: based on the data of \cite{2011arXiv1109.2497M}. To be compared with Fig. 4 of \cite{2019ApJ...874...81F}. Note that one planet with imprecise orbital parameters (and hence bin) is removed from the analysis. \textit{Middle}: based on the data of \cite{2021ApJS..255....8R}. To be compared with Fig. 2 and 3 of \cite{2021ApJS..255...14F}. \textit{Bottom}: same assuming a uniform distribution instead of a beta distribution for the eccentricities.
\label{radial_distribution}} 
\end{figure}

 We use the tools used by \cite{2019ApJ...874...81F} and the orbital parameters found in the present study for planets with periods between 2000 d and 10 yr (which roughly corresponds to \textit{a} < 4 au) in the CH survey, to re-estimate the occurrence rates of LPGP with periods in the 2000d-10yr range, and we fitted the occurrence rates for periods below 10 yr. The turnover mentioned in the CH analysis at 1500-2000 d (Fig. \ref{radial_distribution}, top) is not clearly present any longer. 
 
 Similarly, we use the tools used by \cite{2021ApJS..255...14F} and the orbital parameters revised in the present study to re-estimate the occurrences of LPGP with \textit{a} < 8 au in the CL survey. The LPGPs radial distribution (Fig. \ref{radial_distribution}, middle) no longer shows a turnover at about 2.5 au as found in the CH study, or at about 3.5 au as found in the CL study, but rather a steep increase at about 1 au followed by a flat distribution between 3 and about 8 au. A similar conclusion was tentatively reached by the AAPS study, using albeit different hypotheses for the range of masses considered (see above) and for the planet's orbital properties during the injection-recovery process (circular orbits only). Considering a uniform distribution in eccentricities instead of a beta distribution to compute the completeness leads to similar qualitative conclusions, with yet larger occurrence rates (associated with larger error bars) beyond 1 au (Fig. \ref{radial_distribution}). This latter figure also illustrates the importance of the assumption on the eccentricity distribution. Considering planets with masses in the 1-20 \Mjup range only leads to similar conclusions. Finally, as expected, the estimated occurrence rate of planets for \textit{a} > 1 au is lower (by $\sim$40\%) when considering planets in the 1-20 \Mjup range than when considering planets in the 0.1-20 \Mjup range.

\section{On the detectability of Solar System analogs with RV technique}

Are planetary systems similar to our own common or rare? To quantify the detection capabilities of Jupiter and Saturn analogs using the data gathered by the largest survey published to date, the CL survey, we add the RV signals of planets with Jupiter/Saturn masses, semi-major axes, and eccentricities to the stars RV\footnote{Only stars for which no planet/BD or trends or false positives had been reported are considered.}. Eight phases are considered per planet. The stars are assumed to have different inclinations: 15, 30, 50, and 90 degrees. We fit the RV time series using a genetic algorithm. Planets are regarded as detected if the recovered periods and \msini are within 25\% of the injected ones, and if the recovered eccentricities have an absolute difference of less than 0.1 with respect to the injected values\footnote{No constraint was put on the planet phase as the orbits are almost circular.}. 

The detection rate for Jupiter analogs is found to be 29\% for the most favorably inclined systems (edge-on)\footnote{If no constraint is put on the eccentricity, the rate is 58\% (edge-on) and 47\% (50 degrees).}. For an inclination of 50 degrees, the rate is 18\%. A similar exercise considering Saturn analogs led to detection rates of significantly less than 1\%. These values are optimistic as they do not take into account the impact of stellar activity. 

The present study shows that sub-Jupiter mass planets beyond  $\sim$3-5 au will remain mostly out of reach of RV techniques. As Earth-mass planets orbiting about 1 au from Sun-like stars may also remain out of RV detection capabilities because of the stellar noise (\cite{2019A&A...632A..81M}, \cite{2021MNRAS.505..830C}, \cite{2020A&A...642A.157M}, \cite{2010A&A...512A..39M}), RV techniques alone will probably not be able to find true solar system analogs.

\section{Concluding remarks}
In the present paper, we have analyzed the limitations of fitting incomplete RV time series to characterize the orbital properties and minimum masses of long-period planets. 
We have revisited the orbital and mass solutions of all planets with semi-major axis greater than 3 au (resp. 5 au) reported in the CH (resp. CL) surveys, using, whenever possible, new RV data. This leads to either significantly different solutions or to solutions much less constrained than those reported. 
Then, we discussed the way to estimate the radial distributions of LPGPs. We showed, in particular, the impact of the assumptions on the planets' eccentricity distribution and on the mass distribution of sub-Jupiter mass planets beyond 4 (resp. 8 au). 
 
 This work calls into question the validity of the radial distributions of long-period planets available today, and, in particular, the decrease of the giant planet occurrence rate beyond 3-8 au. It also shows that estimating an accurate radial distribution of giant planets beyond about 5-8 au is not possible with the presently available data, and will remain so at least when considering the entire giant planets’ population (30-6000 $M_{Earth}$). Combining RV and Gaia data is certainly promising for detecting and characterizing Jupiter analogs (\cite{2019A&A...629C...1G}, \cite{2020A&A...642A..31D}, \cite{2021A&A...645A...7K}, \cite{2021AJ....162..266L}, \cite{2022ApJS..262...21F}, \cite{2023A&A...670A..65P}), at least in the case of simple systems closer than 5-10 au (\cite{2018exha.book.....P}, \cite{2018A&A...614A..30R}). Micro-lensing can be sensitive to long-period planets \citep{2016ApJ...833..145S}. The forthcoming Roman Space Telescope will allow us to extend the statistics of LPGPs into the 2-10 au range, with an increased sensitivity for M-type stars compared to earlier type stars (\cite{2021exbi.book....2G}, \cite{2019ApJS..241....3P}). Direct imaging is, in principle, the most suitable for detecting long-period planets. While current systems on ground-based 10m-class telescopes mostly allow for the detection of young and massive (> 5 \Mjupv) giant planets beyond 10 au typically (beyond 5 au in the next decade), and the James Webb Space Telescope will possibly allow for the imaging of Saturn mass planets around close-by late-type stars, extremely large telescopes will be needed to possibly image closer Jupiter and Saturn siblings around mature stars. 

\begin{acknowledgements}
 This project has also received funding from the European Research Council (ERC) under the European Union's Horizon 2020 research and innovation programme (COBREX; grant agreement n° 885593).
 This study was funded by a grant from PSL/OCAV. 
 This publications makes use of the The Data \& Analysis Center for Exoplanets (DACE), which is a facility based at the University of Geneva (CH) dedicated to extrasolar planets data visualisation, exchange and analysis. DACE is a platform of the Swiss National Centre of Competence in Research (NCCR) PlanetS, federating the Swiss expertise in Exoplanet research. The DACE platform is available at https://dace.unige.ch.
 This research has made use of the SIMBAD database and VizieR catalogue access tool, operated at CDS, Strasbourg, France.
 Based on data retrieved from the SOPHIE archive at Observatoire de Haute-Provence (OHP), available at atlas.obs-hp.fr/sophie.
 Based on spectral data retrieved from the ELODIE archive at Observatoire de Haute-Provence (OHP).
 Based on observations collected at the European Southern Observatory under ESO programmes. 
\end{acknowledgements}

\bibliographystyle{aa}
\bibliography{CB.bib}

\onecolumn

\appendix

\section{Long-term stellar activity}

The aim is to estimate the impact of the long-term stellar activity for the 22 CH and 19 CL targets of interest in the present analysis. Because spectroscopic proxies of activity are in general not available, the activity signal of each star is simulated using activity-induced synthetic RV time series representative of old F-G-K main-sequence stars \citep{2019A&A...627A..56M}. These simulations take into account complex activity patterns: i/ a large number of spots and plages, following a solar-like butterfly diagram, and with a large diversity of sizes and lifetime, in agreement with the observations of the Sun, ii/ a small contribution due to granulation and supergranulation \citep{2021MNRAS.505..830C}. The simulations cover a parameter space in terms of variability compatible with observations of each star of interest, based on its averaged log(R’HK). They are typically performed over 1 to 3 activity cycles and for different inclinations chosen randomly between 0° and 90°. 

\begin{figure*}[t!]
 \centering
\includegraphics[width=1\textwidth]{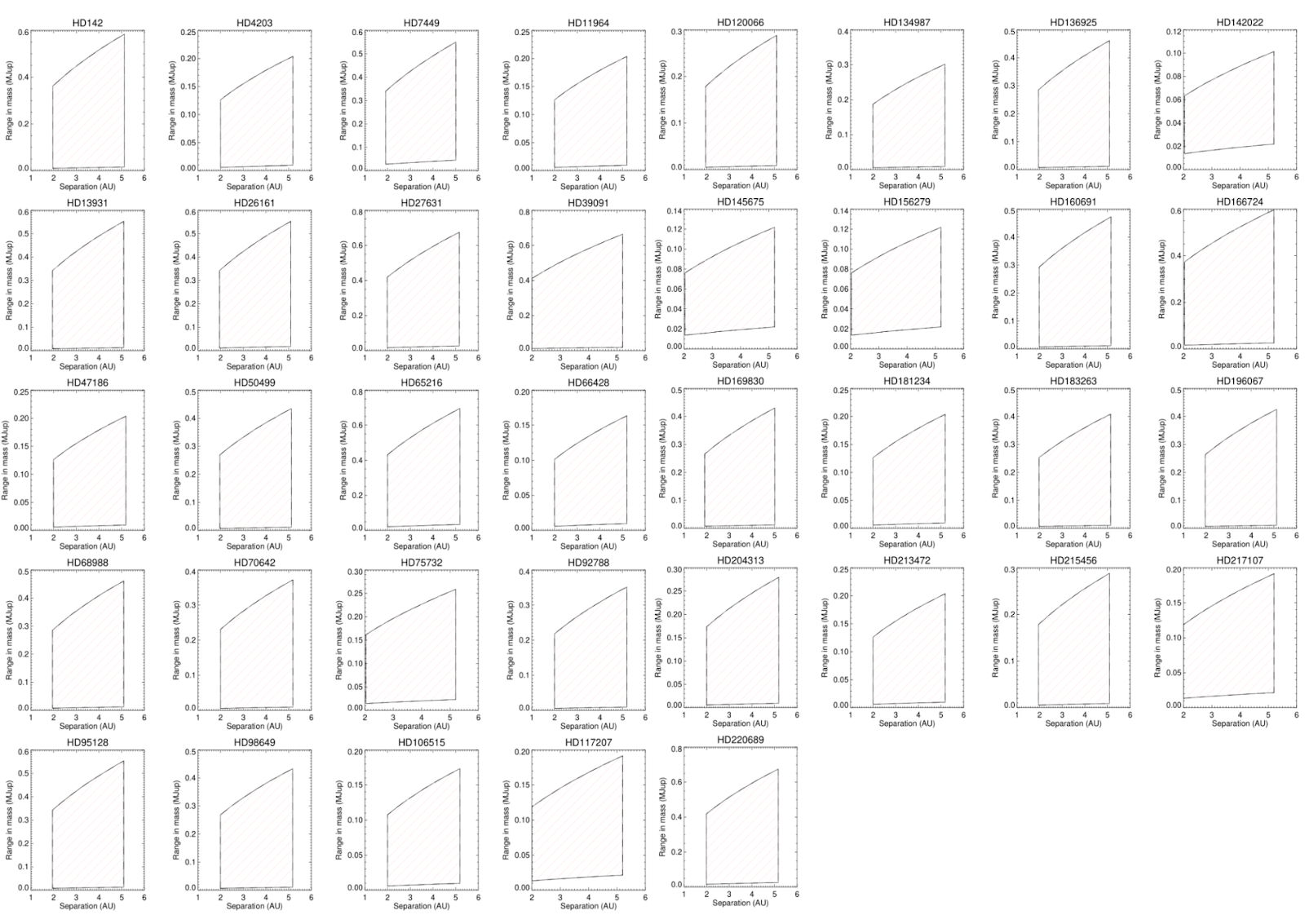}
\caption{Range of a and minimum masses of the planets whose detection/characterization could be significantly impacted by the long-term magnetic cycle of the host star, given the lowest and highest amplitudes of the activity signal deduced from the simulations. The cycle periods are typically between 2.3 and 14.7 yr. 
\label{sma_mass_activity}} 
\end{figure*}

For each target, we select such a simulation corresponding to its spectral type and to an average log(R'HK) within 0.05 of the measured one. The number of simulations depends on the stars but is always greater than 250. For each realization, a random phase is considered. We discard GL 317 because its spectral type (M3) is out of the range of the available simulations. HD 181433 (K5) is also discarded because it is slightly outside the range of the spectral type in the simulation and its average log(R’HK) is much below the lowest activity level in the set of simulations corresponding to the closest spectral type. Finally, for six stars, especially possible subgiants, the observed log(R'HK) is below the minimum value in the simulations. The quietest simulations are therefore considered, and the derived stellar jitter could be slightly overestimated in such cases. 
For each star and each noise realization, the amplitude of the RV variations due to stellar activity over its calendar of observations is computed. The minimum and maximum as well as the average/median amplitudes are then computed, considering all these realizations. The potential impact on the detection of LPGP is estimated and shown in Fig\ref{sma_mass_activity}.

\begin{figure}[t!]
 \centering
\includegraphics[width=0.7\textwidth]{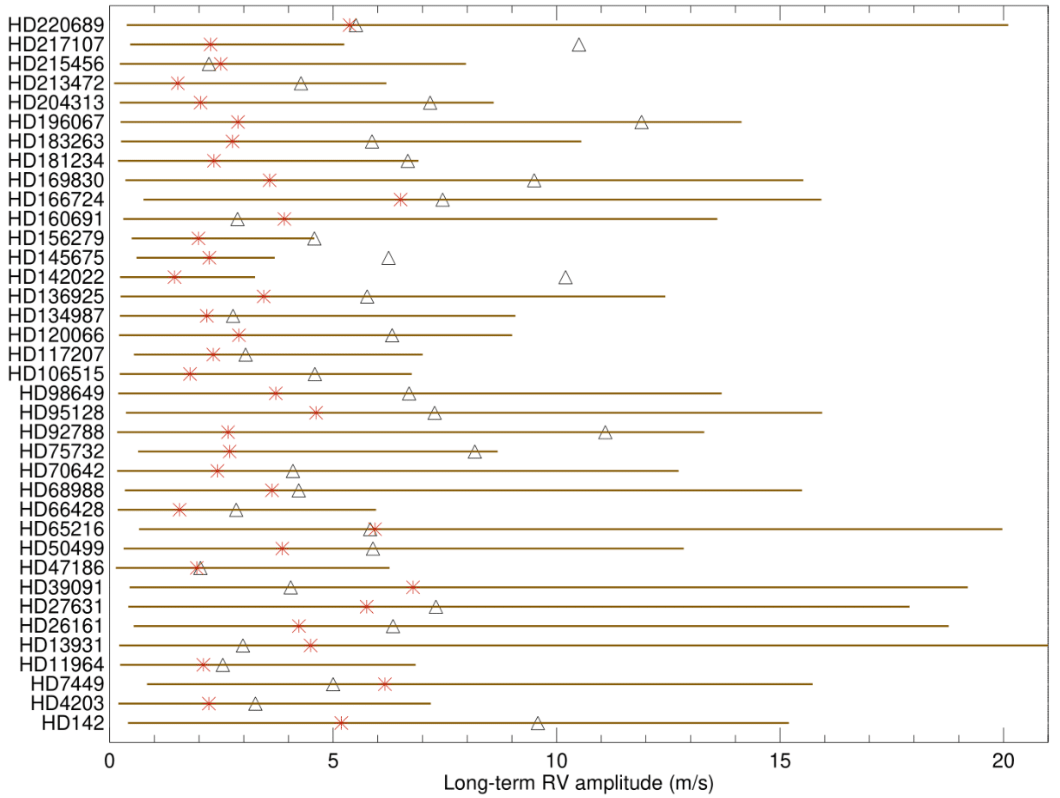}
\caption{Estimated amplitudes of the long-term RV signal due to stellar activity. Triangles indicate the amplitudes measured on the RV curves, and asterisks indicate the median values found by the simulations. 
\label{LPGP_activity}}
\end{figure}

\section{Examples of orbital solutions different from those found in the CH survey analysis}

\subsection{HD 98649b: example of a solution significantly different from the CH survey analysis}

HD 98649 is a 1.03 \Msun, G4 V star \cite{2019A&A...625A..71R}. The CH survey reported an LPGP with a period of 10400 days, a minimum mass of 7 \Mjup, and a high eccentricity of 0.86. Based on 68 RV CORALIE measurements obtained between 2003 and 2019, \cite{2019A&A...625A..71R} reported a period of $6023_{-255}^{+412}$ days, a minimum mass of $6.79_{-0.31}^{+0.53}$ \Mjup, and an eccentricity of $0.86_{-0.02}^{+0.04}$. Recently, combining the CORALIE RV and Hipparcos/Gaia absolute astrometry data, \cite{2021AJ....162..266L} reported an orbital inclination of either $43.7_{-8.1}^{+13}$° or $136.3_{-13}^{+8.1}$°, corresponding to a true mass of $9.7_{-1.9}^{+2.3}$ \Mjupv.

\begin{figure}[t!]
 \centering
\includegraphics[width=0.49\textwidth]{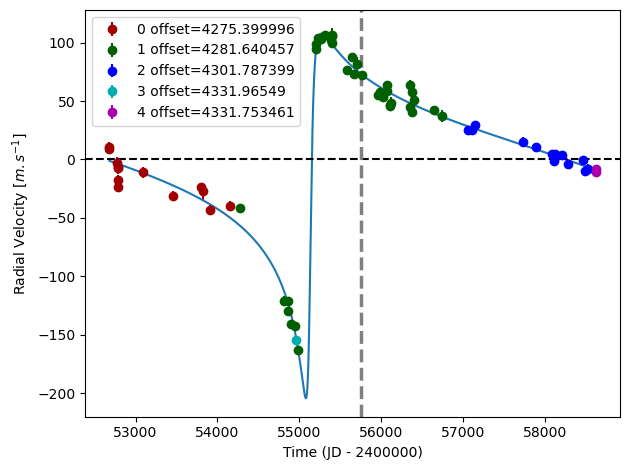}
\includegraphics[width=0.49\textwidth]{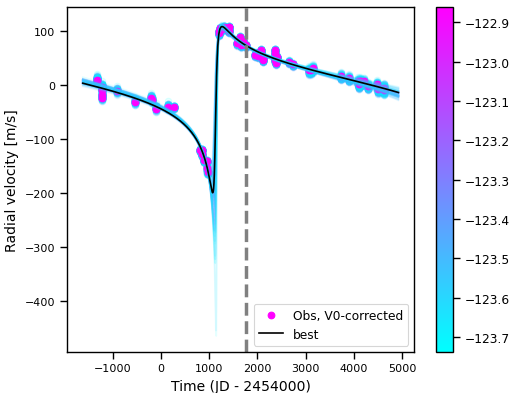}
\caption{Orbital fits for HD 98649b.\textit{Top}: fit of the HD 98649 RV with DPASS. Red - C98; green - C07; blue - C14; cyan - H03; purple - H15. The blue curve shows the best fit. \textit{bottom}: fit of the HD 98649 RV using MCMC. The black curve shows the best fit. The colorbar corresponds to the log-likelihood of the fits. The gray dotted line indicates the end of the CH survey.
\label{fit_HD98649}} 
\end{figure}

In the present study, in addition to the CORALIE dataset, 3 HARPS RV measurements obtained between 2009 and 2019 are considered. DPASS and MCMC (1000 walkers and 400000 iterations) are used to fit the data. We find a period of $5719_{-608}^{+968}$ days, compatible (within the error bars) with that reported for HD 98649b in \cite{2019A&A...625A..71R}. However, for the minimum mass and the eccentricity, larger ranges of solutions are found using MCMC with confidence intervals at 1-$\sigma$ between 6.6 and 9.4 \Mjup for the minimum mass and between 0.84 and 0.95 for the eccentricity. The fits are shown in Fig\ref{fit_HD98649}.

\begin{figure}[t!]
 \centering
\includegraphics[width=1.0\textwidth]{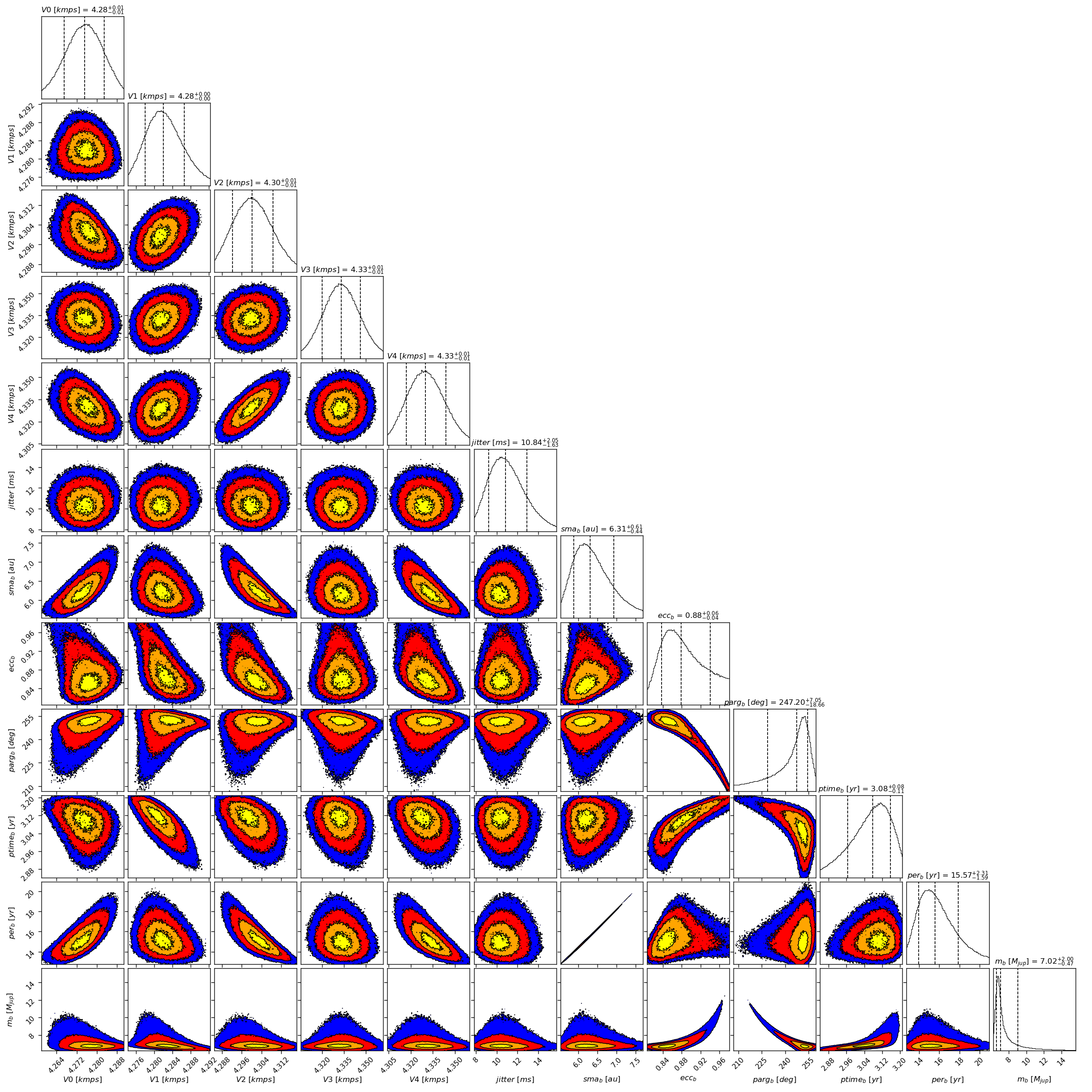}
\caption{Corner plot of posteriors for the one-planet model MCMC fit of HD 98649 RV data.
\label{corner_HD98649}} 
\end{figure}

\subsection{HD 47186c: example of a very poorly constrained solution in the CH survey}

HD 47186 is a 0.99 \Msun, G5 V star \citep{2009A&A...496..527B}. Based on 66 RV HARPS measurements obtained between 2003 and 2008, \cite{2009A&A...496..527B} reported a Hot Jupiter (HD 47186b) signal with a period of $4.0845 \pm 0.0002$ days, a minimum mass 0.07167 \Mjup, and an eccentricity of $0.038 \pm 0.02$ as well as a giant planet (HD 47186c) signal with a period of $1353.6 \pm 57.1$ days, a minimum mass of 0.35 \Mjup, and an eccentricity of $0.249 \pm 0.073$. The CH survey reported properties close to those reported in \cite{2009A&A...496..527B} for HD 47186b, and different properties for HD 47186c, with a period of 3552 days, a minimum mass of 0.58 \Mjup, and an eccentricity of 0.28.

\begin{figure}[t!]
 \centering
\includegraphics[width=0.49\textwidth]{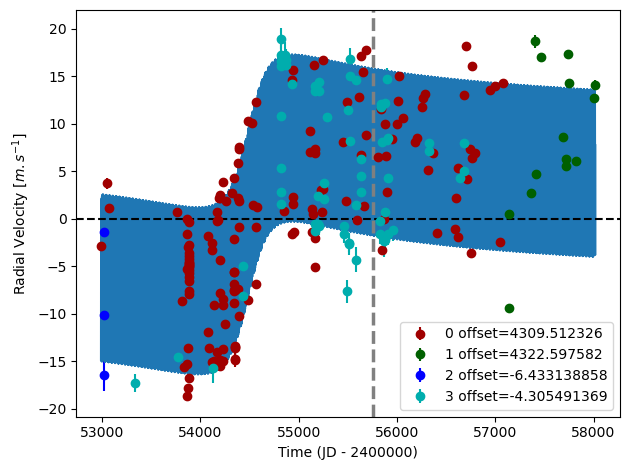}
\includegraphics[width=0.49\textwidth]{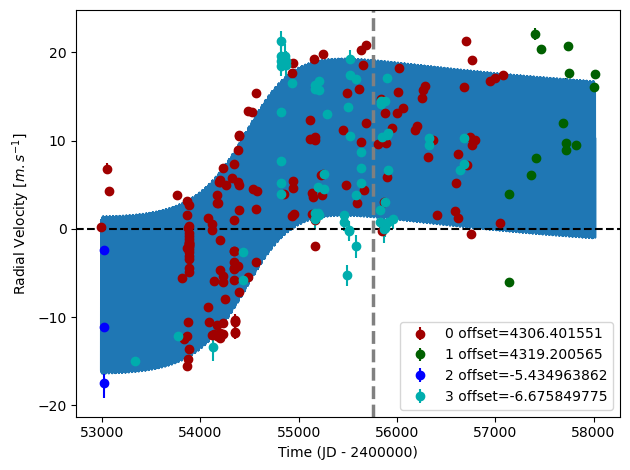}
\includegraphics[width=0.49\textwidth]{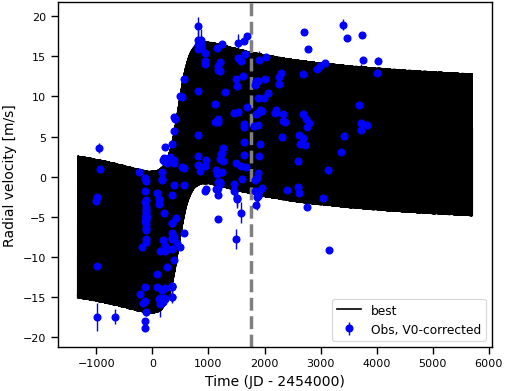}
\caption{Orbital fits for HD 47186 b and c.\textit{Top}: fit of the HD 47186 RV with DPASS. Red - H03; green - H15; blue - Hir94; cyan - Hir04. The blue curve shows the best fit. \textit{Middle}: fit of the HD 47186 RV with DPASS, with the minimum \textit{a} fixed at 100 au. The points are the same as on the \textit{top}. The blue curve shows the best fit. \textit{Bottom}: fit of the HD 47186 RV using MCMC. The black curves show the best fit. The gray dotted line indicates the end of the CH survey.
\label{fit_HD47186}} 
\end{figure}

In the present study, in addition to the dataset used by \cite{2009A&A...496..527B}, we use 105 RV measurements obtained with HARPS between 2003 and 2017 and 67 HIRES RV measurements obtained between 2004 and 2014. While only one minimum of HD 47186c is observed in \cite{2009A&A...496..527B}, a minimum and a maximum are now covered. However, the orbital phase is not yet fully covered. DPASS and MCMC (1000 walkers and 400000 iterations) are used to fit the data. To converge more easily, the priors on the semi-major axis and the minimum mass of the short-period planet HD 47186b are chosen close to the values found in \cite{2009A&A...496..527B}. The properties of planet b are, expectedly, close to those reported in \cite{2009A&A...496..527B}. For HD 47186c, a period of 83984 days, a minimum mass of 0.64 \Mjup, and an eccentricity of 0.93 are found with DPASS, with a corresponding rms of residuals of 3 m/s, and a period between 9790 and 42000 days, a minimum mass of $0.63 \pm 0.05$ \Mjup, and an eccentricity of $0.79_{-0.11}^{+0.09}$ are found using MCMC. The period of HD 47186c is then significantly different from previously published values. It is actually poorly constrained. The fits are shown in Fig \ref{fit_HD47186}.

To explore the range of possible values, the semi-major axis is fixed to different values and the data are fitted with DPASS. \textit{a} up to 100 au do not significantly change the rms of the residuals (3.3 m/s against 3 m/s with a left free). In this case, the minimum mass is 0.98 \Mjup, and the extremely high eccentricity is 0.95, corresponding to the maximum value of the eccentricity prior. As the RV curve of HD 47186c covers a maximum and a minimum, the RV offset is well-constrained and changing it will not significantly change the solution.

The properties found in the CH survey for HD47186c are then not confirmed. Moreover, the available RV data do not allow us to properly determined the orbital parameters and the minimum mass of the LPGP. Additional data are needed to further constrain its orbital properties.

\begin{figure}[t!]
 \centering
\includegraphics[width=1.0\textwidth]{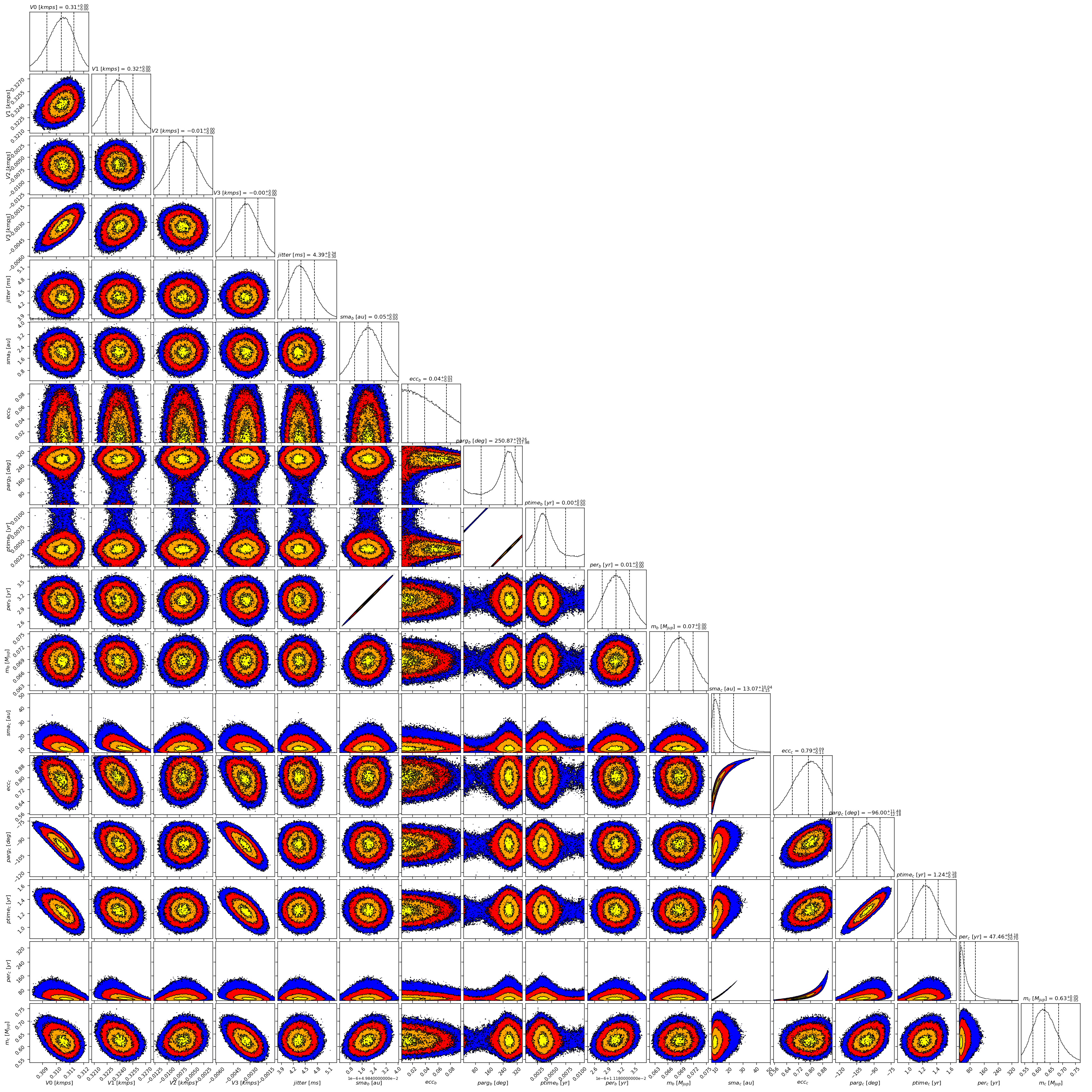}
\caption{Corner plot of posteriors for the two-planets model MCMC fit of HD 47186 RV data.
\label{corner_HD47186}} 
\end{figure}

\section{Corner plots of the cases described in section 3.2}

\begin{figure}[t!]
 \centering
\includegraphics[width=1.0\textwidth]{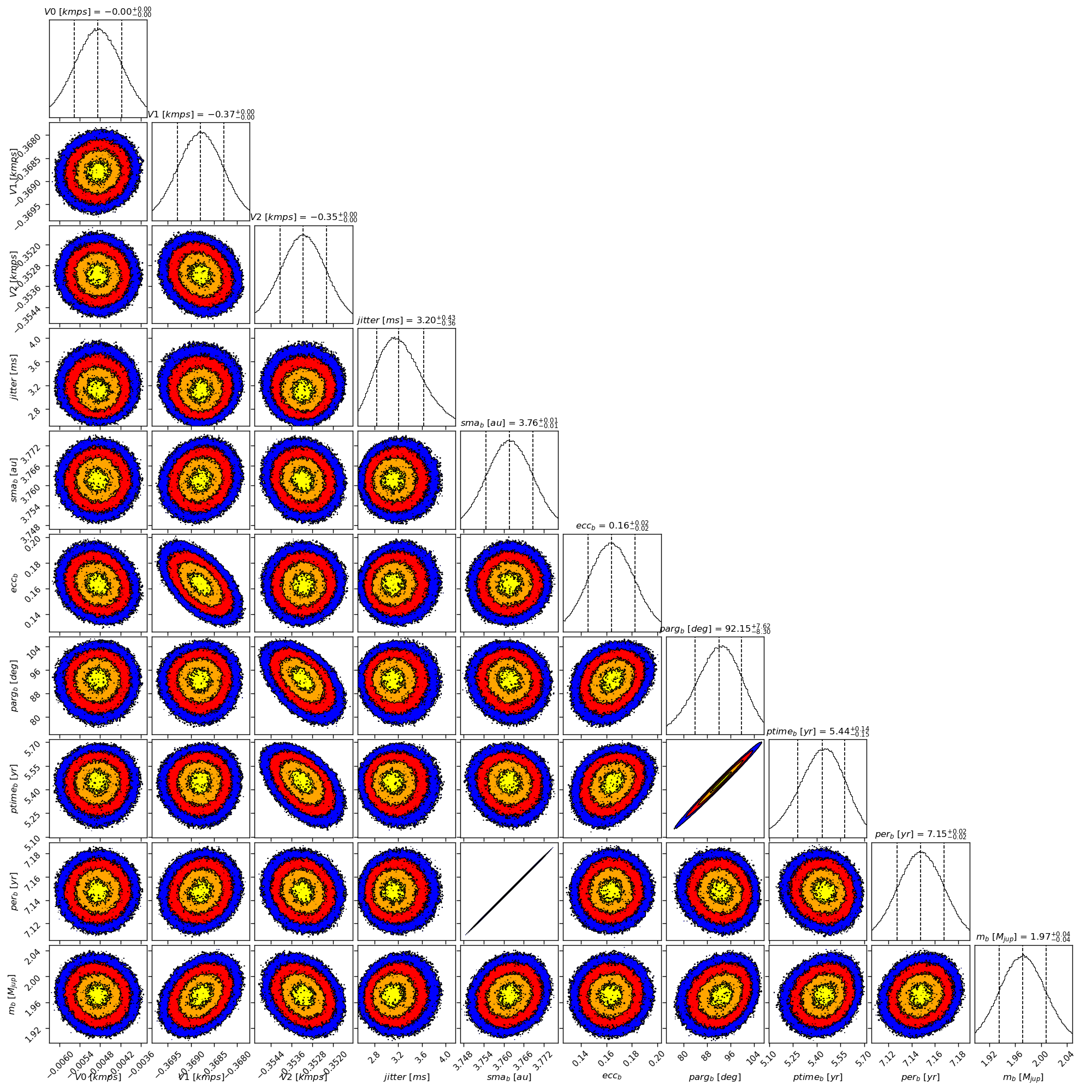}
\caption{Corner plot of posteriors for the two-planets model MCMC fit of HD 117207 RV data.
\label{corner_HD117207}} 
\end{figure}

\begin{figure}[t!]
 \centering
\includegraphics[width=1.0\textwidth]{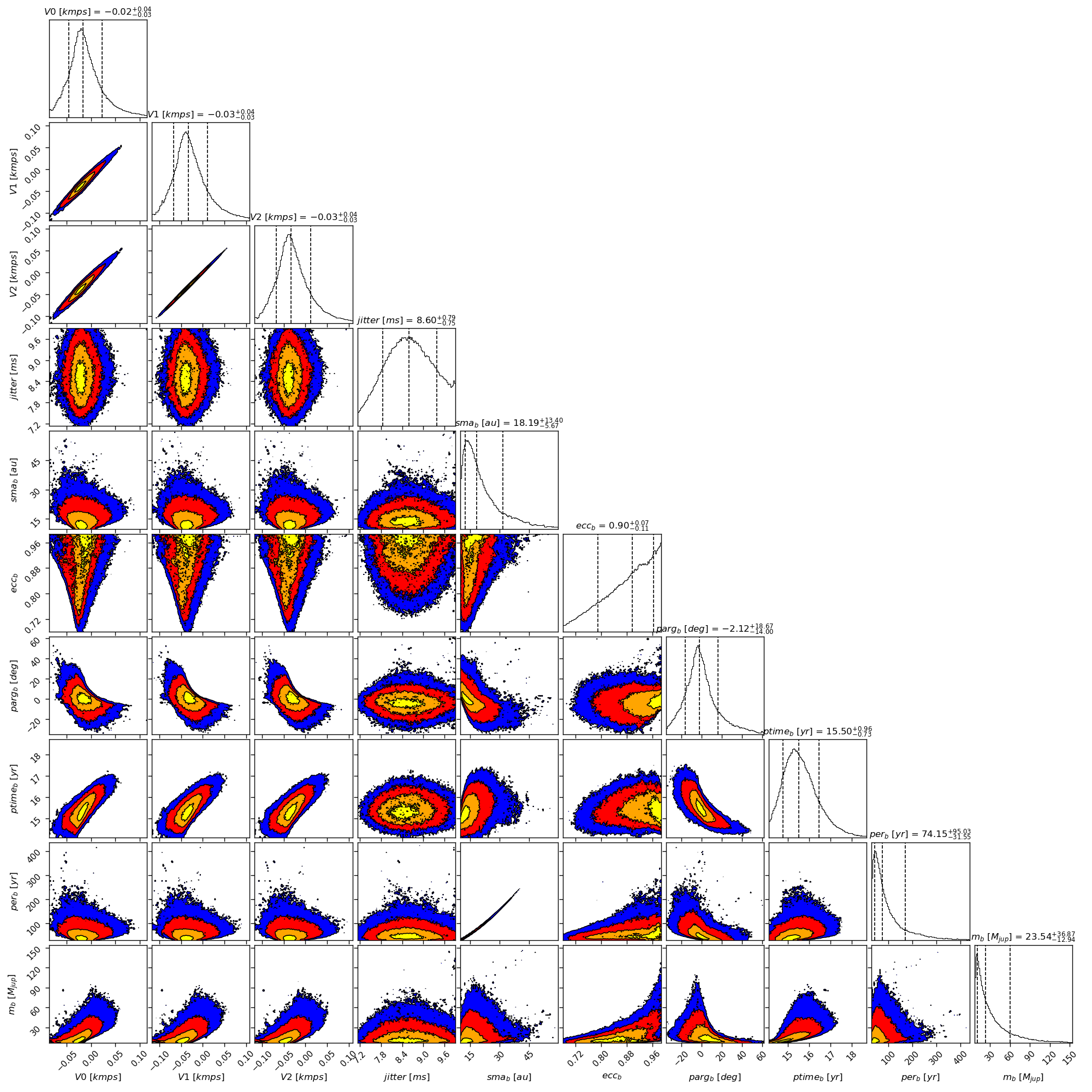}
\caption{Corner plot of posteriors for the one-planet model MCMC fit of HD 26161 RV data with loose priors on the RV offsets and on semi-major axes.
\label{corner_HD 26161_loose_priors}} 
\end{figure}

\begin{figure}[t!]
 \centering
\includegraphics[width=1.0\textwidth]{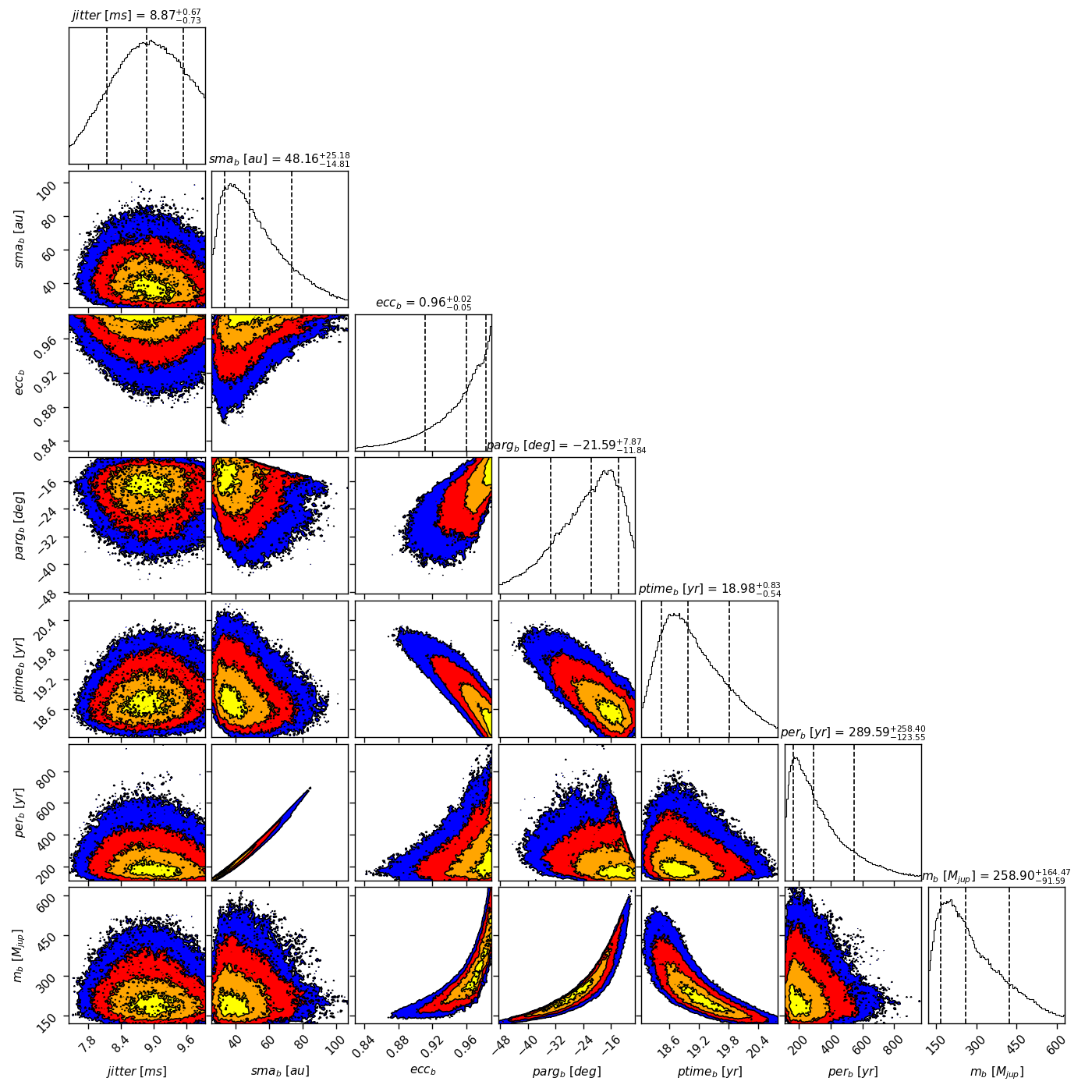}
\caption{Corner plot of posteriors for the one-planet model MCMC fit of HD 26161 RV data with the RV offset fixed at 300m/s.
\label{corner_HD 26161_offset}} 
\end{figure}

\begin{figure}[t!]
 \centering
\includegraphics[width=1.0\textwidth]{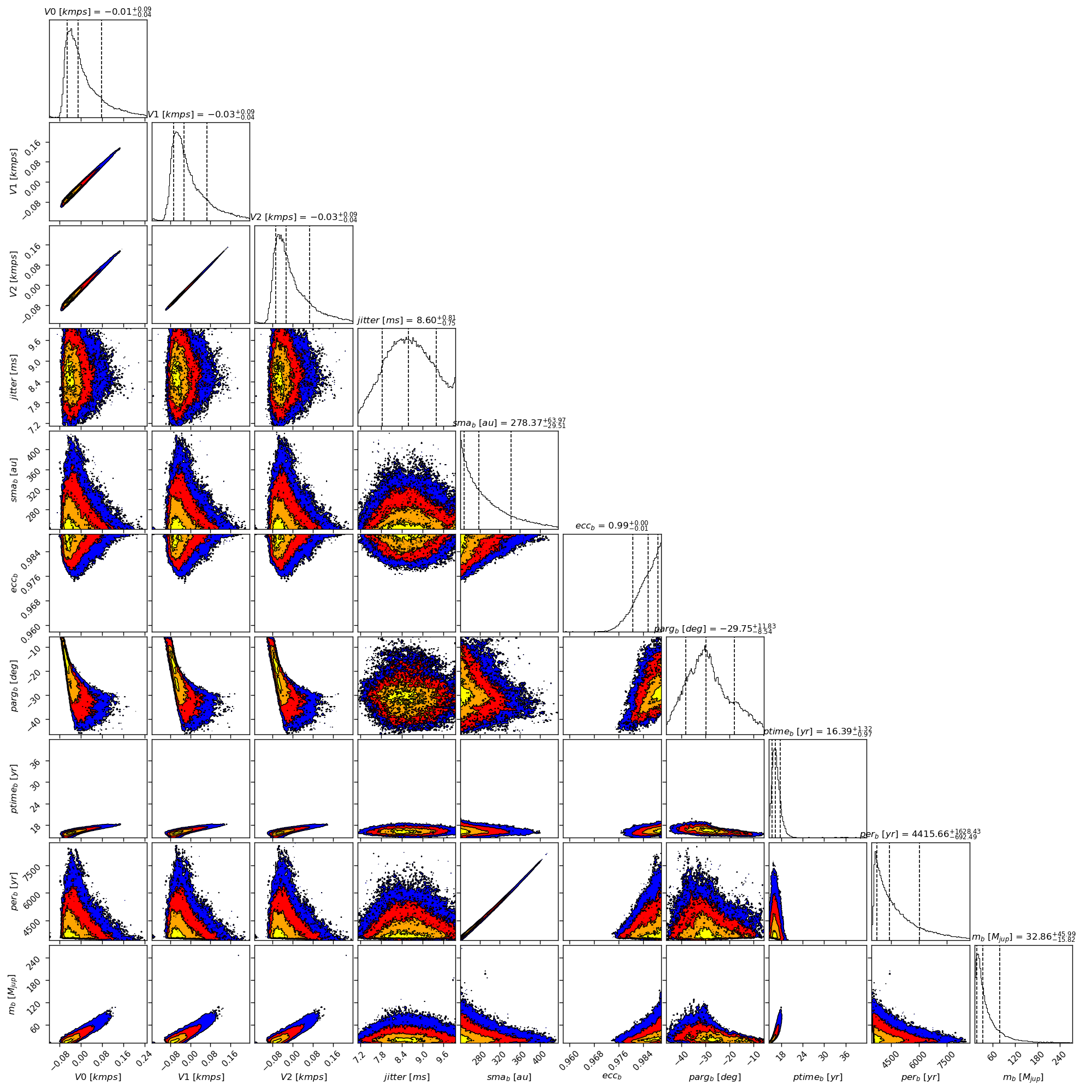}
\caption{Corner plot of posteriors for the one-planet model MCMC fit of HD 26161 RV data with a prior on a in the range 240-300au.
\label{corner_HD 26161_sma}} 
\end{figure}

\begin{figure}[t!]
 \centering
\includegraphics[width=1.0\textwidth]{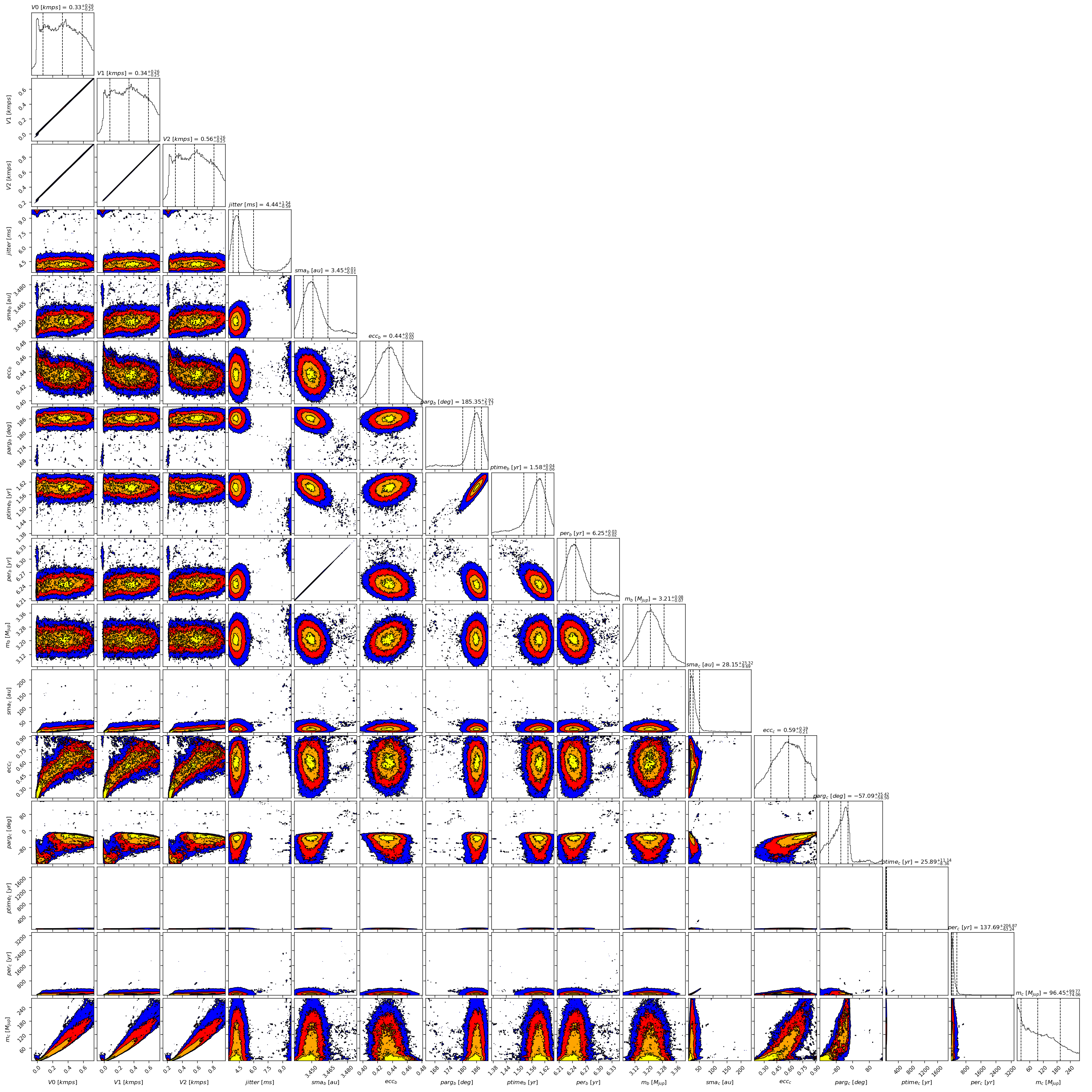}
\caption{Corner plot of posteriors for the two-planets model MCMC fit of HD 66428 RV data.
\label{corner_HD66428}} 
\end{figure}

\section{Summary tables}

\begin{table*}[h!]
\centering
\caption{Comparison of the orbital parameters and masses of the CH survey with those obtained in this study for planets with orbital parameters and/or minimum mass poorly constraints or significantly different from those found in the CH survey.}
\begin{adjustbox}{width=\textwidth}
\begin{tabular}[h!]{cccccccccccc}
\hline
 Star & \multicolumn{2}{c}{Period (days)} & \multicolumn{2}{c}{\textit{a} (AU)} & \multicolumn{2}{c}{\msini (\Mjupv)} & \multicolumn{2}{c}{Eccentricity} & \multicolumn{2}{c}{Baseline (days)} \\ 
\hline
 & CH survey & This paper & CH survey & This paper & CH survey & This paper & CH survey & This paper & CH survey & This paper \\
\hline
 HD 142 & b = 350.3 & \underline{DPASS} & b = 1.0 & \underline{DPASS} & b = 1.31 & \underline{DPASS} & b = 0.25 & \underline{DPASS} & 1359 & 7860 \\
 & & b = 351.4 & & b = 1.02 & & b = 1.1 & & b = 0.15 & & \\
 & & c = 10016 & & c = 9.6 & & c = 10.4 & & c = 0.27 & & \\
 & & \underline{MCMC} & & \underline{MCMC} & & \underline{MCMC} & & \underline{MCMC} & & \\
 & & b = $351.5_{-0.7}^{+0.6}$ & & b = $1.02 \pm 0.1$ & & b = $1.14_{-0.07}^{+0.06}$ & & b = $0.15_{-0.07}^{+0.06}$ & & \\
 & & c = $10111_{-493}^{+716}$ & & c = $9.6_{-0.3}^{+0.4}$ & & c = $10.4 \pm 0.5$ & & c = $0.27 \pm 0.03$ & & \\
\hline
HD 7449 & b = 1275 & \underline{DPASS} & b = 2.3 & \underline{DPASS} & b = 1.11 & \underline{DPASS} & b = 0.82 & \underline{DPASS} & 4029 & 5903 \\
 & c = 4046 & b = 1263 & c = 5.1 & b = 2.3 & c = 2 & b = 0.82 & c = 0.53 & b = 0.65 & & \\
 & & c = 7845 & & c = 7.9 & & c = 6.4 & & c = 0.23 & & \\
 & & \underline{MCMC} & & \underline{MCMC} & & \underline{MCMC} & & \underline{MCMC} & & \\
 & & b = $1256 \pm 11$ & & b = $2.32_{-0.02}^{+0.01}$ & & b = $0.84 \pm 0.07$ & & b = $0.66 \pm 0.05$ & & \\
 & & c = 19500 - 39700 & & c = 15 - 24 & & c = 35 - 189 & & c < 0.36 & & \\
\hline
HD 47186 & b = 4.08 & \underline{DPASS} & b = 0.05 & \underline{DPASS} & b = 0.072 & \underline{DPASS} & b = 0.04 & \underline{DPASS} & 2698 & 5029 \\
 & c = 3552 & b = 4.08 & c = 4.5 & b = 0.05 & c = 0.58 & b = 0.073 & c = 0.28 & b = 0.65 & & \\
 & & c = 83984 (UP 366900) & & c = 37.4 (UP 100) & & c = 0.64 (UP 0.98) & & c = 0.93 (UP 0.95) & & \\
 & & \underline{MCMC} & & \underline{MCMC} & & \underline{MCMC} & & \underline{MCMC} & & \\
 & & b = $4.085 \pm 0.001$ & & b = $0.050 \pm 0.001$ & & b = $0.069_{-0.004}^{+0.003}$ & & b < 0.07 & & \\
 & & c = 9790 - 42000 & & c = 9 - 24 & & c = $0.63 \pm 0.05$ & & c = $0.79_{-0.11}^{+0.09}$ & & \\
\hline
 HD 50499 & b = 2457.87 & \underline{DPASS} & b = 3.9 & \underline{DPASS} & b = 1.74 & \underline{DPASS} & b = 0.25 & \underline{DPASS} & 4480 & 8439 \\
 & & b = 2470 & & b = 3.92 & & b = 1.42 & & b = 0.32 & & \\
 & & c = 9000 (UP 320000) & & c = 9.3 (UP 100) & & c = 2.7 (UP 14) & & c = 0.18 (UP 0.92) & & \\
 & & \underline{MCMC} & & \underline{MCMC} & & \underline{MCMC} & & \underline{MCMC} & & \\
 & & b = $2470 \pm 15$ & & b = $3.91_{-0.01}^{+0.02}$ & & b = $1.45 \pm 0.08$ & & b = $0.31_{-0.03}^{+0.04}$ & & \\
 & & c = $8500 - 19700$ & & c = $8.9 - 15.7$ & & c = $2.6 - 4.6$ & & c = $0.14 - 0.35$ & & \\
\hline
HD 65216 & b = 579 & \underline{DPASS} & b = 1.3 & \underline{DPASS} & b = 1.41 & \underline{DPASS} & b = 0.26 & \underline{DPASS} & 4039 & 5371 \\
 & c = 5542 & b = 577 & c = 6 & b = 1.3 & c = 2.24 & b = 1.3 & c = 0.15 & b = 0.17 & & \\
 & & c = 5381 (UP 117800) & & c = 5.8 (UP 45) & & c = 2 (UP 2.2) & & c = 0.3 (UP 0.9) & & \\
 & & \underline{MCMC} & & \underline{MCMC} & & \underline{MCMC} & & \underline{MCMC} & & \\
 & & b = $591_{-7}^{+14}$ & & b = $1.30_{-0.01}^{+0.02}$ & & b = $1.24_{-0.08}^{+0.09}$ & & b = $0.29_{-0.07}^{+0.06}$ & & \\
 & & c = 5800 - 28000 & & c = 6 - 17 & & c = 1.7 - 3.5 & & c = 0.23 - 0.73 & & \\
\hline
HD 98649 & b = 10400 & \underline{DPASS} & b = 9.4 & \underline{DPASS} & b = 7 & \underline{DPASS} & b = 0.86 & \underline{DPASS} & 3024 & 5943 \\
 & & b = 5628 & & b = 6.3 & & b = 7 & & b = 0.87 & & \\
 & & \underline{MCMC} & & \underline{MCMC} & & \underline{MCMC} & & \underline{MCMC} & & \\
 & & b = $5719_{-608}^{+968}$ & & b = $6.3_{-0.4}^{+0.7}$ & & b = $6.6 - 9.4$ & & b = 0.84 - 0.95 & & \\
\hline
HD 134987 & b = 258.18 & \underline{DPASS} & b = 0.8 & \underline{DPASS} & b = 1.56 & \underline{DPASS} & b = 0.23 & \underline{DPASS} & 2275 & 7228 \\
 & c = 5000 & b = 258.2 & c = 5.9 & b = 0.81 & c = 0.8 & b = 1.6 & c = 0.11 & b = 0.22 & & \\
 & & c = 6316 & & c = 6.8 & & c = 0.99 & & c = 0.0 & & \\
 & & \underline{MCMC} & & \underline{MCMC} & & \underline{MCMC} & & \underline{MCMC} & & \\
 & & b = $258.40 \pm 0.01$ & & b = $0.81 \pm 0.01$ & & b = $1.61 \pm 0.02$ & & b = $0.23 \pm 0.01$ & & \\
 & & c = $6315_{-263}^{+276}$ & & c = $6.8_{-0.2}^{+0.3}$ & & c = $0.98_{-0.05}^{+0.06}$ & & c < 0.17 & & \\
\hline
HD 142022 & b = 1928 & \underline{DPASS} & b = 3 & \underline{DPASS} & b = 4.47 & \underline{DPASS} & b = 0.52 & \underline{DPASS} & 3968 & 3968 \\
 & & b = 1940 & & b = 3.04 & & b = 4.5 (UP 35) & & c = 0.51 (UP 0.95) & & \\
 & & \underline{MCMC} & & \underline{MCMC} & & \underline{MCMC} & & \underline{MCMC} & & \\
 & & b = $1987_{-44}^{+55}$ & & b = $3.04_{-0.05}^{+0.06}$ & & b = $4 - 24$ & & b > 0.47 & & \\
\hline
$\mu$ Ara & b = 9.64 & \underline{DPASS} & b = 0.091 & \underline{DPASS} & b = 0.033 & \underline{DPASS} & b = 0.12 & \underline{DPASS} & 4456 & 6342 \\
 (HD 160691) & c = 313.2 & b = 9.6 & c = 0.9 & b = 0.09 & c = 0.6 & b = 0.03 & c = 0.04 & b = 0.09 & & \\
 & d = 648.7 & c = 308 & d = 1.5 & c = 0.9 & d = 1.72 & c = 0.4 & d = 0.18 & c = 0.06 & & \\
 & e = 8723 & d = 645 & e = 8.5 & d = 1.5 & e = 2.49 & d = 1.6 & e = 0.43 & d = 0.0 & & \\
 & & e = 3965 & & e = 5 & & e = 1.8 & & e = 0.08 & & \\
 & & \underline{MCMC} & & \underline{MCMC} & & \underline{MCMC} & & \underline{MCMC} & & \\
 & & b = $9.64 \pm 0.01$ & & b = $0.091 \pm 0.001$ & & b = $0.033_{-0.001}^{+0.002}$ & & b < 0.10 & & \\
 & & c = $309.0 \pm 0.1$ & & c = $0.92 \pm 0.01$ & & c = $0.43_{-0.01}^{+0.02}$ & & c = $0.04_{-0.01}^{+0.02}$ & & \\
 & & d = $645.0 \pm 0.4$ & & d = $1.50 \pm 0.01$ & & d = $1.63 \pm 0.01$ & & d = $0.05_{-0.01}^{+0.02}$ & & \\
 & & e = $3944_{-36}^{+39}$ & & e = $5.02_{-0.04}^{+0.03}$ & & e = $1.84 \pm 0.03$ & & e = $0.07 \pm 0.01$ & & \\
\hline
HD 166724 & b = 8100 & \underline{DPASS} & b = 7.4 & \underline{DPASS} & b = 4.12 & \underline{DPASS} & b = 0.77 & \underline{DPASS} & 3507 & 4018 \\
 & & b = 4880 (UP 60000) & & b = 5.3 (UP 28) & & b = 3.5 & & c = 0.74 (UP 0.95) & & \\
 & & \underline{MCMC} & & \underline{MCMC} & & \underline{MCMC} & & \underline{MCMC} & & \\
 & & b = 4500 - 6400 & & b = 5.0 - 6.3 & & b = $3.5 \pm 0.2$ & & b = $0.74_{-0.03}^{+0.04}$ & & \\
\hline
HD 181433 & b = 9.37 & \underline{DPASS} & b = 0.08 & \underline{DPASS} & b = 0.023 & \underline{DPASS} & b = 0.42 & \underline{DPASS} & 2737 & 5026 \\
 & c = 1019 & b = 9.37 & c = 1.8 & b = 0.08 & c = 0.7 & b = 0.02 & c = 0.25 & b = 0.35 & & \\
 & d = 3201 & c = 1018 & d = 3.9 & c = 1.82 & d = 0.58 & c = 0.68 & d = 0.11 & c = 0.24 & & \\
 & & d = 6847 (UP 413290) & & d = 6.5 (UP 100) & & d = 0.6 (UP 0.91) & & d = 0.48 (UP 0.95) & & \\
 & & \underline{MCMC} & & \underline{MCMC} & & \underline{MCMC} & & \underline{MCMC} & & \\
 & & b = $9.37 \pm 0.01$ & & b = $0.080 \pm 0.001$ & & b = $0.023_{-0.003}^{+0.002}$ & & b = $0.33 \pm 0.09$ & & \\
 & & c = $1022_{-3}^{+4}$ & & c = $1.82 \pm 0.01$ & & c = $0.67_{-0.01}^{+0.02}$ & & c = $0.24 \pm 0.02$ & & \\
 & & d = $6000 - 9600$ & & d = $6.0 - 8.2$ & & d = $0.60 \pm 0.02$ & & d = $0.50_{-0.07}^{+0.10}$ & & \\
\hline
HD 196067 & b = 4100 & \underline{DPASS} & b = 5 & \underline{DPASS} & b = 7.1 & \underline{DPASS} & b = 0.63 & \underline{DPASS} & 4372 & 7478 \\
 & & b = 3508 (UP 7259) & & b = 4.9 (UP 8) & & b = 6.6 & & c = 0.6 (UP 0.68) & & \\
 & & \underline{MCMC} & & \underline{MCMC} & & \underline{MCMC} & & \underline{MCMC} & & \\
 & & 1) b = $3835_{-220}^{+263}$ & & 1) b = $5.0 \pm 0.2$ & & 1) b = 5.8 - 13.5 & & 1) b = 0.57 - 0.86 & & \\
 & & 2) b = 5977 - 8100 & & 2) b = 7.0 - 8.6 & & 2) b = 5.7 - 10.1 & & 2) b = 0.66 - 0.88 & & \\
\hline
HD 217107 & b = 7.12 & \underline{DPASS} & b = 0.073 & \underline{DPASS} & b = 1.4 & \underline{DPASS} & b = 0.12 & \underline{DPASS} & 471 & 7412 \\
 & c = 4270 & b = 7.1 & c = 5.2 & b = 0.073 & c = 2.62 & b = 1.3 & c = 0.51 & b = 0.13 & & \\
 & & c = 5135 & & c = 5.9 & & c = 4 & & c = 0.41 & & \\
 & & \underline{MCMC} & & \underline{MCMC} & & \underline{MCMC} & & \underline{MCMC} & & \\
 & & b = $7.13 \pm 0.01$ & & b = $0.075 \pm 0.001$ & & b = $1.37 \pm 0.01$ & & b = $0.12 \pm 0.01$ & & \\
 & & c = $5128_{-52}^{+49}$ & & c = $6.00 \pm 0.04$ & & c = $4.2 \pm 0.2$ & & c = $0.42 \pm 0.03$ & & \\
\hline
\end{tabular}
\end{adjustbox}
\textbf{Notes :} The present study includes the results obtained with a genetic algorithm (DPASS) and those obtained with an MCMC. With the MCMC, 68\% confidence intervals are given for each parameter and the median is only given when the probability distribution has a profile close to a Gaussian distribution. When the orbits are poorly constrained, the maximum values of \textit{a} and offsets considered are indicated (UP), as well as the corresponding posteriors of the fits.
\label{PMvalues}
\end{table*}

\begin{table*}[h!]
\centering
\caption{Comparison of the orbital parameters and masses of the CL survey with those obtained in this study for planets with orbital parameters and/or minimum mass poorly constraints.}
\begin{adjustbox}{width=\textwidth}
\begin{tabular}[h!]{ccccccccccc}
\hline
 Star & \multicolumn{2}{c}{Period (days)} & \multicolumn{2}{c}{\textit{a} (AU)} & \multicolumn{2}{c}{\msini (\Mjupv)} & \multicolumn{2}{c}{Eccentricity} & Baseline (days) \\ 
\hline
 & CL survey & This paper & CL survey & This paper & CL survey & This paper & CL survey & This paper & \\
\hline
HD 4203 & b = $437_{-0.2}^{+0.1}$ & \underline{DPASS} & b = $1.177_{-0.022}^{+0.021}$ & \underline{DPASS} & b = $1.821_{-0.077}^{+0.078}$ & \underline{DPASS} & b = $0.513_{-0.014}^{+0.013}$ & \underline{DPASS} & 7124 \\
 & c = $7424.7_{-1064.9}^{+8949.9}$ & b = 437.1 & c = $7.8_{-0.78}^{+5.4}$ & b = 1.17 & c = $2.68_{-0.24}^{+0.99}$ & b = 1.8 & c = $0.19_{-0.089}^{+0.29}$ & b = 0.51 & \\
 & & c = 6642 (UP 969370) & & c = 7.2 (UP 200) & & c = 2.5 (UP 6.4) & & c = 0.16 (UP 0.95) & \\
 & & \underline{MCMC} & & \underline{MCMC} & & \underline{MCMC} & & \underline{MCMC} & \\
 & & b = $437.1 \pm 0.2$ & & b = $1.17 \pm 0.01$ & & b = $1.82 \pm 0.06$ & & b = $0.51 \pm 0.02$ & \\
 & & c = $7500 - 40200$ & & c = $7.2 - 24$ & & c = $2.5 - 5.1$ & & c = $0.18 - 0.65$ & \\
\hline
HD 26161 & b = $32000_{-10000}^{+21000}$ & \underline{DPASS} & b = $20.4_{-4.9}^{+7.9}$ & \underline{DPASS} & b = $13.5_{-3.7}^{+8.5}$ & \underline{DPASS} & b = $0.82_{-0.05}^{0.061}$ & \underline{DPASS} & 8076 \\
 & & b = 64224 (UP 1265126) & & b = 33 (UP 240) & & b = 14.1 (UP 80) & & b = 0.78 (UP 0.95) & \\
 & & \underline{MCMC} & & \underline{MCMC} & & \underline{MCMC} & & \underline{MCMC} & \\
 & & b = $20000 - 155000$ & & b = $15 - 60$ & & b = $13 - 170$ & & b > 0.75 & \\
\hline
 HD 50499 & b = $2461.7_{-15.3}^{+15.9}$ & \underline{DPASS} & b = $3.847_{-0.04}^{+0.038}$ & \underline{DPASS} & b = $1.346_{-0.087}^{+0.084}$ & \underline{DPASS} & b = $0.348_{-0.045}^{+0.046}$ & \underline{DPASS} & 8439 \\
 & c = $10419.8_{—1264.7}^{+3220.8}$ & b = 2470 & c = $10.1_{-0.84}^{+2}$ & b = 3.92 & c = $3.18_{-0.46}^{+0.63}$ & b = 1.42 & c = $0.241_{-0.075}^{+0.089}$ & b = 0.32 & \\
 & & c = 9000 (UP 320000) & & c = 9.3 (UP 100) & & c = 2.7 (UP 14) & & c = 0.18 (UP 0.92) & \\
 & & \underline{MCMC} & & \underline{MCMC} & & \underline{MCMC} & & \underline{MCMC} & \\
 & & b = $2470 \pm 15$ & & b = $3.91_{-0.01}^{+0.02}$ & & b = $1.45 \pm 0.08$ & & b = $0.31_{-0.03}^{+0.04}$ & \\
 & & c = $8500 - 19700$ & & c = $8.9 - 15.7$ & & c = $2.6 - 4.6$ & & c = $0.14 - 0.35$ & \\
\hline
 HD 66428 & b = $2288.9_{-6.8}^{+6.1}$ & \underline{DPASS} & b = $3.455_{-0.05}^{+0.049}$ & \underline{DPASS} & b = $3.19 \pm 0.11$ & \underline{DPASS} & b = $0.418_{-0.014}^{+0.015}$ & \underline{DPASS} & 6937 \\
 & c = $39000_{—18000}^{+56000}$ & b = 2273 & c = $23_{-7.6}^{+19}$ & b = 3.92 & c = $27_{-17}^{+22}$ & b = 3.19 & c = $0.418_{-0.014}^{+0.015}$ & b = 0.45 & \\
 & & c = 9107 (UP 931850) & & c = 8.7 (UP 190) & & c = 2.2 (UP 52) & & c = 0.18 (UP 0.95) & \\
 & & \underline{MCMC} & & \underline{MCMC} & & \underline{MCMC} & & \underline{MCMC} & \\
 & & b = $2281_{-7}^{+10}$ & & b = $3.45 \pm 0.01$ & & b = $3.21 \pm 0.07$ & & b = $0.44_{-0.02}^{+0.01}$ & \\
 & & c = $27000 - 125000$ & & c = $18 - 51$ & & c = $22 - 196$ & & c = $0.38 - 0.78$ & \\
\hline
 HD 68988 & b = $6.27642 \pm 0.00001$ & \underline{DPASS} & b = $0.0702 \pm 0.0001$ & \underline{DPASS} & b = $1.915_{-0.054}^{+0.053}$ & \underline{DPASS} & b = $0.1581_{-0.0031}^{+0.0027}$ & \underline{DPASS} & 7354 \\
 & c = $16100_{—3500}^{+11000}$ & b = 6.28 & c = $13.2_{-2.0}^{+5.3}$ & b = 0.07 & c = $15_{-1.5}^{+2.8}$ & b = 1.96 & c = $0.45_{-0.081}^{+0.13}$ & b = 0.15 & \\
 & & c = 12190 (UP 108000) & & c = 11.1 (UP 47.5) & & c = 13.9 (UP 28.6) & & c = 0.34 (UP 0.8) & \\
 & & \underline{MCMC} & & \underline{MCMC} & & \underline{MCMC} & & \underline{MCMC} & \\
 & & b = $6.281 \pm 0.001$ & & b = $0.071 \pm 0.001$ & & b = $1.96 \pm 0.01$ & & b = $0.15 \pm 0.01$ & \\
 & & c = $13200 - 44800$ & & c = $11.6 - 26$ & & c = $14 - 21$ & & c = $0.37 - 0.68$ & \\
\hline
 HD 92788 & b = $332.39 \pm 0.53$ & \underline{DPASS} & b = $0.949 \pm 0.013$ & \underline{DPASS} & b = $3.52 \pm 0.10$ & \underline{DPASS} & b = $0.3552_{-0.0072}^{+0.0070}$ & \underline{DPASS} & 8076 \\
 & c = $8356.2_{—387.2}^{+535.0}$ & b = 326 & c = $8.26_{-0.28}^{+0.37}$ & b = 0.95 & c = $2.81_{-0.17}^{+0.18}$ & b = 3.5 & c = $0.355_{-0.052}^{+0.057}$ & b = 0.36 & \\
 & & c = 8000 (UP 710000) & & c = 8 (UP 160) & & c = 2.9 (UP 4.2) & & c = 0.33 (UP 0.95) & \\
 & & \underline{MCMC} & & \underline{MCMC} & & \underline{MCMC} & & \underline{MCMC} & \\
 & & b = $325.70_{-0.03}^{+0.04}$ & & b = $0.95 \pm 0.01$ & & b = $3.50 \pm 0.05$ & & b = $0.36 \pm 0.01$ & \\
 & & c = $8167_{-448}^{+950}$ & & c = $8.2_{-0.4}^{+0.6}$ & & c = $2.9 \pm 0.3$ & & c = $0.35_{-0.06}^{+0.07}$ & \\
\hline
 HD 95128 & b = $1076.6_{-1.1}^{+1.3}$ & \underline{DPASS} & b = $2.059_{-0.033}^{+0.031}$ & \underline{DPASS} & b = $2.438_{-0.085}^{+0.086}$ & \underline{DPASS} & b = $0.0160_{-0.0080}^{+0.0076}$ & \underline{DPASS} & 11957 \\
 (47UMa) & c = $2286.5_{-16.4}^{+18.8}$ & b = 1075 & c = $3.403_{-0.056}^{0.055}$ & b = 2.1 & c = $0.497_{-0.030}^{+0.032}$ & b = 2.4 & c = $0.179_{-0.092}^{+0.090}$ & b = 0.05 & \\
 & d = $18656.8_{-4005.5}^{+10542.6}$ & c = 2266 & d = $13.8_{-2.1}^{+4.8}$ & c = 3.4 & d = $1.51_{-0.18}^{+0.22}$ & c = 0.5 & d = $0.38_{-0.15}^{+0.16}$ & c = 0.3 & \\
 & & d = 12963 (UP 12114000) & & d = 10.9 (UP 590) & & d = 1.4 (UP 173) & & d = 0.32 (UP 0.95) & \\
 & & \underline{MCMC} & & \underline{MCMC} & & \underline{MCMC} & & \underline{MCMC} & \\
 & & b = $1079_{-4}^{+1}$ & & b = $2.08 \pm 0.01$ & & b = $2.44 \pm 0.04$ & & b = $0.04 \pm 0.01$ & \\
 & & c = $2294 \pm 22$ & & c = $3.48 \pm 0.03$ & & c = $0.50_{-0.03}^{+0.04}$ & & c = $0.14_{-0.10}^{+0.13}$ & \\
 & & d = $150000 - 410000$ & & d = $40 - 113$ & & d = $2 - 148$ & & d = $0.39 - 0.91$ & \\
\hline
 HD 120066 & b = $42500.9_{-19485.3}^{+56701.6}$ & \underline{DPASS} & b = $25_{-8.3}^{+19}$ & \underline{DPASS} & b = $3.15_{-0.17}^{+0.18}$ & \underline{DPASS} & b = $0.886_{-0.056}^{+0.049}$ & \underline{DPASS} & 8504 \\
 & & b = 32435 (UP 206500) & & b = 20.4 (UP 70) & & b = 3.3 (UP 3.9) & & b = 0.86 (UP 0.95) & \\
 & & \underline{MCMC} & & \underline{MCMC} & & \underline{MCMC} & & \underline{MCMC} & \\
 & & b = $18400 - 78300$ & & b = $14 - 37$ & & b = $3.3 \pm 0.1$ & & b = $0.86_{-0.07}^{+0.06}$ & \\
\hline
 HD 145675 & b = $1766.4 \pm 0.07$ & \underline{DPASS} & b = $2.830_{-0.041}^{+0.040}$ & \underline{DPASS} & b = $4.85_{-0.14}^{+0.15}$ & \underline{DPASS} & b = $0.3674_{-0.0038}^{+0.0035}$ & \underline{DPASS} & 8739 \\
 (14Her) & c = $25000_{—9200}^{+24000}$ & b = 1767 & c = $16.4_{-4.3}^{+9.3}$ & b = 2.8 & c = $5.8_{-1.0}^{+1.4}$ & b = 4.6 & c = $0.3674_{-0.035}^{+0.038}$ & b = 0.36 & \\
 & & c = 11421 (UP 760000) & & c = 9.6 (UP 110) & & c = 3.3 (UP 13.2) & & c = 0.19 (UP 0.92) & \\
 & & \underline{MCMC} & & \underline{MCMC} & & \underline{MCMC} & & \underline{MCMC} & \\
 & & b = $1767 \pm 1$ & & b = $2.77 \pm 0.01$ & & b = $4.66 \pm 0.04$ & & b = $0.36 \pm 0.01$ & \\
 & & c = $18500 - 100200$ & & c = $13 - 41$ & & c = $4.5 - 9.8$ & & c = $0.32 - 0.75$ & \\
\hline
 HD 213472 & b = $16700_{-4800}^{+12000}$ & \underline{DPASS} & b = $13_{-2.6}^{+5.7}$ & \underline{DPASS} & b = $3.48_{-0.59}^{+1.10}$ & \underline{DPASS} & b = $0.530_{-0.085}^{+0.120}$ & \underline{DPASS} & 6761 \\
 & & b = 17609 (UP 1006075) & & b = 13.5 (UP 200) & & b = 4 (UP 700) & & b = 0.08 (UP 0.92) & \\
 & & \underline{MCMC} & & \underline{MCMC} & & \underline{MCMC} & & \underline{MCMC} & \\
 & & b = $55000 - 814000$ & & b = $29 - 193$ & & b = $84 - 700$ & & b = $0.43 - 0.81$ & \\
\hline
 GL 317 & b = $722.33 \pm 0.37$ & \underline{DPASS} & b = $1.1799 \pm 0.0076$ & \underline{DPASS} & b = $1.852 \pm 0.037$ & \underline{DPASS} & b = $0.098 \pm 0.016$ & \underline{DPASS} & 7302 \\
 & c = $7831.2_{—721.1}^{+1469.1}$ & b = 695.2 & c = $5.78_{-0.38}^{+0.74}$ & b = 1.15 & c = $1.673_{-0.075}^{+0.078}$ & b = 1.79 & c = $0.098 \pm 0.016$ & b = 0.08 & \\
 & & c = 7351 (UP 31074) & & c = 5.5 (UP 14.5) & & c = 1.6 (UP 1.75) & & c = 0.22 (UP 0.95) & \\
 & & \underline{MCMC} & & \underline{MCMC} & & \underline{MCMC} & & \underline{MCMC} & \\
 & & b = $695_{-0.5}^{+0.7}$ & & b = $1.15 \pm 0.01$ & & b = $1.79 \pm 0.03$ & & b = $0.09_{-0.02}^{+0.01}$ & \\
 & & c = $6970 - 11500$ & & c = $5.3 - 6.9$ & & c = $1.6 \pm 0.1$ & & c = $0.17 - 0.41$ & \\
\hline
\end{tabular}
\end{adjustbox}
\textbf{Notes :} The present study includes the results obtained with a genetic algorithm (DPASS) and those obtained with an MCMC. With the MCMC, 68\% confidence intervals are given for each parameter and the median is only given when the probability distribution has a profile close to a Gaussian distribution. When the orbits are poorly constrained, the maximum values of \textit{a} and offsets considered are indicated (UP), as well as the corresponding posteriors of the fits.
\label{PMvalues}
\end{table*}

\end{document}